\documentclass[aps,nofootinbib,superscriptaddress,onecolumn,12pt,tightenlines,floatfix,showkeys]{revtex4-2}

\usepackage{intrinsic}
\usepackage{hyperref}

\begin{document}
\ifproofpre{}{\count\footins = 1000}


\title{Intrinsic operators for the translationally-invariant many-body problem}

\author{Mark A. Caprio}
\affiliation{Department of Physics, University of Notre Dame, Notre Dame, Indiana 46556-5670, USA}

\author{Anna E. McCoy}
\affiliation{Department of Physics, University of Notre Dame,
Notre Dame, Indiana 46556-5670, USA}
\affiliation{TRIUMF, Vancouver, British Columbia V6T 2A3, Canada}

\author{Patrick J. Fasano}
\affiliation{Department of Physics, University of Notre Dame, Notre Dame, Indiana 46556-5670, USA}

\date{\today}

\begin{abstract}
%
%
The need to enforce fermionic antisymmetry in the nuclear many-body problem
commonly requires use of single-particle coordinates, defined relative to some
fixed origin.
%
%
To obtain physical operators which nonetheless act on the nuclear many-body
system in a Galilean-invariant fashion, thereby avoiding spurious center-of-mass
contributions to observables, it is necessary to express these operators with
respect to the translational intrinsic frame.
%
%
Several commonly-encountered operators in nuclear many-body calculations,
including the magnetic dipole and electric quadrupole operators (in the impulse
approximation), and generators of $\grpu{3}$ and $\grpsptr$ symmetry groups, are
bilinear in the coordinates and momenta of the nucleons and, when expressed in
intrinsic form, become two-body operators.  To work with such operators in a
second-quantized many-body calculation, it is necessary to relate three distinct
forms: the defining intrinsic-frame expression, an explicitly two-body
expression in terms of two-particle relative coordinates, and a decomposition
into one-body and separable two-body parts.
%
%
We establish the relations between these forms, for general (non-scalar and
non-isoscalar) operators bilinear in coordinates and momenta.
%
%
\relax
%
%
\relax
%
%
\relax
\end{abstract}



\maketitle



\section{Introduction}
\label{sec-intro}

In studying the nuclear system, the states of interest are those involving
excitation of the intrinsic structure of the nucleus in its comoving inertial
reference frame, not those involving ``spurious'' motion of the center of mass.
The nuclear many-body problem is translationally invariant.  Except for the
contribution to the kinetic energy operator arising from center-of-mass motion,
the problem is, moreover, Galilean invariant, \textit{i.e.}, also invariant under boosts to the
momenta.  Therefore, if the nuclear problem could be solved in the full,
untruncated many-body space, then the spectrum of nuclear excited states would
contain an intrinsic spectrum, reflecting intrinsic excitations of the nucleons
relative to each other.  Onto this
intrinsic spectrum would be superposed an infinite continuum of copies, each
representing the combination of this intrinsic structure with a different
center-of-mass motion.

In practical calculations, there is a fundamental conflict between choosing a
computational framework which manifestly reflects the Galilean invariant nature
of the Hamiltonian and one which readily respects the antisymmetry constraints
imposed by the fermionic statistics of the nucleons.  A natural starting point
for the translationally invariant problem is provided by a coordinate system,
such as Jacobi coordinates, which explicitly separates the center-of-mass
coordinate from the other, intrinsic
coordinates~\cite{navratil1999:spurious-faddeev,navratil2009:ncsm}.  However,
defining a many-body basis in terms of the intrinsic coordinates is challenging.
In particular, it becomes prohibitively difficult to impose antisymmetry in
Jacobi coordinates~\cite{barnea1997:hyperspherical-harmonics-symmetrized} as the
number of particles increases.

Antisymmetry is more easily enforced by representing the system in terms of
antisymmetrized products (Slater determinants) of single-particle states.  These
single-particle states are defined with respect to a common, fixed origin, that
is, with respect to laboratory-frame coordinates.  Thus, the many-body problem
is typically solved in a framework which does not manifestly preserve
translational invariance.

Even so, it is possible to compute observables as they would be measured in the
intrinsic frame, insensitive to the center-of-mass motion.  To do so, we must
ensure that we calculate observables using suitable Galilean-invariant intrinsic
operators, which reduce to the operator of interest when evaluated in the
intrinsic
frame~\cite{gartenhaus1957:com,lipkin1958:com-shell,seyfferth1967:intrinsic-nuclear-part1-algebra,*seyfferth1967:intrinsic-nuclear-part2-commutation,*seyfferth1967:intrinsic-nuclear-part3-oscillator,eisenberg1988:v2,mosconi1987:cm-siegert}.

Many operators of interest for the nuclear many-body problem involve
angular-momentum coupled products of the form $(\xvec \times \xvec)$, $(\xvec
\times \pvec)$, and $(\pvec \times \pvec)$, that is, bilinear in coordinates
and/or momenta.  Although these operators are one-body operators when expressed
in the laboratory frame, they become two-body operators when realized as
intrinsic operators.  To work with the Galilean-invariant intrinsic operators
obtained from such bilinear expressions, in a second-quantized many-body
calculation, we therefore need to evaluate their two-body matrix elements.  To
do so, it is necessary to relate three distinct representations of these
operators: (1)~the defining intrinsic-frame expression, (2)~an explicitly
two-body expression in terms of two-particle relative coordinates, and (3)~a
decomposition into one-body and separable two-body parts.

In the case of the rotational scalar intrinsic kinetic energy operator, commonly
used in nuclear configuration interaction calculations, the relations among
these forms are long familiar to shell model
practitioners~\cite{brussaard1977:shell,glaudemans1985:light-nuclei-isnsm84}.
The structure of the intrinsic squared radius
operator~\cite{bacca2012:6he-hyperspherical,caprio2012:csbasis,caprio2014:cshalo},
which enters into nuclear charge radius calculations (and furthermore serves as
the leading-order electric monopole operator), is essentially identical to that
of the kinetic energy, but with the introduction of charge or isosopin
dependence.  Among the higher-multipolarity electromagnetic transition
operators~\cite{gartenhaus1957:com,eisenberg1988:v2}, in the impulse
approximation, the magnetic dipole and electric quadrupole operators are
likewise bilinear in coordinates and/or momenta, though no longer scalar
operators like those just mentioned.

Bilinear operators, beyond representing physical observables, serve to define
the symmetry structure of the nuclear many-body problem.  The full set of
bilinears provide the generators of the symplectic group $\grpsptr$ in three
dimensions~\cite{rosensteel1977:sp6r-shell,rosensteel1980:sp6r-shell,rowe1985:micro-collective-sp6r}.
This group is closely linked to the dynamics of the many-body problem in
three-dimensional space and contains Elliott's $\grpu{3}$
group~\cite{elliott1958:su3-part1,*elliott1958:su3-part2,*elliott1963:su3-part3,*elliott1968:su3-part4,harvey1968:su3-shell}
as a subgroup.  The intrinsic forms of all these operators share a common
structure.

Here we derive systematic extensions of the relations for intrinsic operators,
from the familiar rotational scalar and isoscalar case, to the general case of
rotational nonscalar and isovector bilinear operators.  These results are
motivated for use in multiple contexts, including in calculating intrinsic
observables in the \textit{ab initio} no-core configuration interaction (NCCI) [or
  no-core shell model (NCSM)] approach~\cite{barrett2013:ncsm} and in
establishing the calculational machinery for the symplectic no-core
configuration interaction (SpNCCI)
framework~\cite{mccoy2018:diss,mccoy2018:spncci-busteni17,mccoyxxxx:spfamilies}.
While the derivations are straightforward, in principle, careful attention must
be paid to the various phase and normalization factors which arise if the
results are to be practically useful in nuclear many-body calculations.  These
include $A$-dependent (or $N$- and $Z$-dependent) counting factors, as well as
insidious factors of $2$ which can easily be overlooked by the unwary when
relating operators on the two-body relative system and the full $A$-body system.

The basic ideas and approaches developed here may be applied further to
operators defined in terms of higher-order products of the coordinates and/or
momenta.  For instance, the electric multipole operator of rank $\lambda$ may
be expressed as an angular-momentum coupled product $(\xvec \times \xvec \times
\cdots \times \xvec)$ of order $\lambda$ in the coordinates.  Although this
operator is a one-body operator when expressed in the laboratory frame, the
corresponding intrinsic operator is a $\lambda$-body operator.


For motivation and context, we first provide some elaboration of the ways in
which the relations considered here enter into both traditional and
symmetry-adapted NCCI nuclear many-body calculations
(Sec.~\ref{sec-background}).  There are ample opportunities for ambiguity
arising from alternative conventions for translating between relative and
single-particle coordinates.  We therefore next carefully set out notation and
definitions for one-body and two-body operators, relative and center-of-mass
coordinates for the two-body system, and intrinsic coordinates for the many-body
system (Sec.~\ref{sec-formalism}).

Before addressing the bilinear intrinsic operators, it is instructive to first
consider the electric dipole operator as an intrinsic operator
(Sec.~\ref{sec-dipole}).  Although this operator is simply linear (not bilinear)
in coordinates, it provides a more transparent context in which to establish the
approaches required for the bilinear operators.

We finally turn to the generic nonscalar bilinear operator, in both its
isoscalar and isovector variants.  We lay out how such an operator is
represented as a two-body intrinsic operator, and then how it may be represented
in terms of either separable or relative two-body operators
(Sec.~\ref{sec-bilinear}).  Preliminary results were presented in
Ref.~\cite{mccoy2018:diss}.

In appendices, we provide additional, more specific results for practical
reference in working with two-body, bilinear intrinsic operators: We review the
explicit expressions for several physically relevant operators, cast in the
generic bilinear form (Appendix~\ref{sec-app-physical}).  We account for the
effect of the proton-neutron mass difference on the
intrinsic kinetic energy operator
(Appendix~\ref{sec-app-isovector-kinetic}).  We note the summation identities
needed in converting between one-body and two-body forms of operators
(Appendix~\ref{sec-app-sum}).  We give expressions casting the one-body spin
operators as two-body operators, so that they can be included in calculations of
two-body matrix elements for the intrinsic magnetic dipole operator
(Appendix~\ref{sec-app-spin}).  We obtain relations for evaluating
isospin-reduced matrix elements of the isovector forms of the two-body intrinsic
operators (Appendix~\ref{sec-app-isospin}).  Then, to facilitate work with the
harmonic oscillator creation and annihilation (ladder) operators, we review definitions and relations for
the harmonic oscillator ladder operators on the
single-particle, relative, and intrinsic degrees of freedom
(Appendix~\ref{sec-app-ladder}).


\section{Background: Intrinsic operators in the nuclear many-body problem}
\label{sec-background}

The Galilean-invariant intrinsic form, which does not introduce center-of-mass
contamination, is obtained, for an operator expressed in coordinates $\xvec$
and/or momenta $\pvec$, by the substitutions $\xvec_i\rightarrow\xvec'_i$ and
$\pvec_i\rightarrow\pvec'_i$.  The Galilean-invariant intrinsic coordinates
$\xvec'_i$ and momenta $\pvec'_i$ are defined relative to the center of mass and
center of momentum, respectively.
As context for the results for general bilinear operators obtained below, we
review some essential observations for the intrinsic kinetic
energy~\cite{eisenberg1976:v3,brussaard1977:shell,glaudemans1985:light-nuclei-isnsm84}
and squared radius operators (Sec.~\ref{sec-background-obs}), then comment on
the implications of the existence of intrinsic $\grpsptr\supset\grpu{3}$ group
generators for the structure of the space used in many-body calculations
(Sec.~\ref{sec-background-space}).  The special properties of the $\Nmax$
truncation~\cite{barrett2013:ncsm} for NCCI calculations follow as a special
case.

\subsection{Kinetic energy and observables}
\label{sec-background-obs}

The ``naive'' physical Hamiltonian for the nuclear many-body
system, $H=T+V$, as obtained in the laboratory frame, involves the total kinetic
energy $T=(2m_N)^{-1}\sum_i \pvec_i\cdot\pvec_i$ of the
nucleons.\footnote{\label{footnote-pn-mass}Here, for simplicity, the same mass
  $m_N$ is taken for all nucleons, in practice commonly defined as $m_N =
  \tfrac{1}{2} (m_p + m_n)$, but more properly taken as $m_N =(Zm_p +
  Nm_n)/A$~\cite{kamuntavicius1999:isoscalar-hamiltonians,gueorguiev2010:nuclear-mass-a-body-interaction-rila10}.
  The full intrinsic kinetic energy operator, with explicit dependence on the
  nucleon masses, is treated in Appendix~\ref{sec-app-isovector-kinetic}.}
While the many-body interaction operator $V$ is Galilean invariant, the kinetic
energy operator violates Galilean invariance, in that the total (center-of-mass)
momentum of the system enters into the kinetic energy.  The full kinetic energy is thus not
invariant under Galilean (momentum) boosts.

Galilean invariance is recovered by substituting the intrinsic
momenta $\pvec'_i$ for the momenta $\pvec_i$, giving the intrinsic kinetic energy
\begin{equation}
  \label{eqn-Tp-sum-kinetic}
    T'
  =\frac{1}{2m_N}\sum_i
  \Bigl(\pvec_i-        \frac1A\sum_j\pvec_j\Bigr)
  \cdot
  \Bigl(\pvec_i-        \frac1A\sum_k\pvec_k\Bigr).
\end{equation}
Several useful observations and relations may be obtained, more or less
directly, from this expression.

First, the kinetic energy then separates as $T=T'+\Tcm$, into intrinsic and
center-of-mass contributions~\cite{bethe1937:nuclear-kinetic-com}.
Equivalently, the intrinsic kinetic energy $T'=T-\Tcm$ may thus be considered to
have had the center-of-mass kinetic energy ``subtracted out'', as
\begin{equation}
  \label{eqn-Tp-separated-kinetic}    
  T'
  = \frac{1}{2m_N} \underbrace{
  \sum_i \pvec_i\cdot\pvec_i }_{\text{One-body}} - \frac1{2Am_N} \underbrace{
      \bigl(\sum_{i}\pvec_i\bigr) \cdot \bigl( \sum_{j}\pvec_j\bigr)
    }_{\text{Center-of-mass}}.
\end{equation}

Then, while the lab-frame $T$ is a one-body operator, the intrinsic $T'$ is a
two-body operator.  It is related, by an $A$-dependent counting factor, to the
sum of two-nucleon relative kinetic energies, as
\begin{equation}
\label{eqn-Tp-tbo-sum-combined-kinetic}    
T'=
\frac{1}{4 m_N A}
\underbrace{
\sumprime_{ij}
\bigl(\pvec_i-\pvec_j\bigr)
\cdot
\bigl(\pvec_i-\pvec_j\bigr)
}_{\mathclap{\text{Relative two-body}}}
 ,
\end{equation}
where the primed sum omits diagonal ($i=j$) terms.
It may therefore be characterized as a relative two-body operator.  The
essential input for working with a two-body operator in a many-body calculation
is its two-body matrix elements.  The explicit relative two-body
form~(\ref{eqn-Tp-tbo-sum-combined-kinetic}) is well-suited for work with a
harmonic oscillator basis.  If the matrix elements are evaluated in an
oscillator basis on the relative coordinate, then the full two-body matrix
elements (between antisymmetrized products of harmonic oscillator
single-particle wave functions) are readily obtained through the Moshinsky
transformation~\cite{brody1960:moshinsky,moshinsky1996:oscillator}.

Finally, $T'$ may be decomposed into one-body and two-body separable contributions as
\begin{equation}
\label{eqn-Tp-obo-separable-tbo-sum-kinetic}    
T'=
\frac{1}{2m_N}\Bigl(1-\frac{1}{A}\Bigr)
\underbrace{
  \sum_i \pvec_i\cdot\pvec_i
}_{\text{One-body}}
-
\frac{1}{2m_N A}
\underbrace{
\sumprime_{ij} \pvec_i\cdot\pvec_j
  }_{\mathclap{\text{Two-body separable}}}.
\end{equation}
The one-body matrix elements of the one-body term are readily evaluated as
radial integrals, while two-body matrix elements of a separable operator are
readily evaluated as products of one-body matrix elements via Racah's reduction
formula~\cite{brink1994:am} (see discussions for the intrinsic kinetic energy,
in particular, in
Refs.~\cite{glaudemans1985:light-nuclei-isnsm84,caprio2012:csbasis}).  This
approach makes no assumptions as to the form of the radial wave functions for
the single-particle states, and thus its relevance is not confined to the
harmonic oscillator basis.

To illustrate calculation of an intrinsic observable, let us take the
root-mean-square (r.m.s.) radius of the nucleus.  The expectation value
\begin{math}
  \tme{\Psi}{\,(\sum_i \xvec_i\cdot\xvec_i)\,}{\Psi},
\end{math}
yields the summed squared radius of the
nucleon probability distribution in the laboratory frame, that is, around the
arbitrary coordinate origin introduced in defining the many-body problem.  A
laboratory-frame r.m.s.\ radius is then obtained as
$r=[A^{-1}\tme{\Psi}{\,(\sum_i \xvec_i\cdot\xvec_i)\,}{\Psi} ]^{1/2}$.  Since the operator
entering into the expectation value is a one-body operator, this expectation
value is readily calculated in second-quantized formalism.

The nucleon density distribution in the laboratory frame, however, reflects both
the intrinsic distribution of nucleons relative to the nuclear center of mass
and the excursions of this center of mass relative to the origin.  (Indeed, if
the motion of the nucleons cleanly factorizes into intrinsic and
center-of-mass parts, then the laboratory-frame density is simply the
convolution of the nucleon density in the translational intrinsic frame and the
probability density of the center-of-mass wave function with respect to the
center-of-mass coordinate.)

The r.m.s.\ radius accessed in experiment is the intrinsic observable, arising
only from the intrinsic structure relative to the center of mass.  This
intrinsic radius is instead obtained from an expectation value defined in terms
of intrinsic coordinates as $r'=[A^{-1}\tme{\Psi}{\,(\sum_i
    \xvec'_i\cdot\xvec'_i)\,}{\Psi} ]^{1/2}$.  This expectation value now
involves a two-body operator, of the same bilinear form as the intrinsic kinetic
energy above.  Relations directly analogous
to~(\ref{eqn-Tp-sum-kinetic})--(\ref{eqn-Tp-obo-separable-tbo-sum-kinetic})
above apply to the intrinsic radius, along with the corresponding observations
about evaluating two-body matrix elements.  Although we have taken the density
of the full nucleon distribution here, including both protons and neutrons,
analogous arguments apply to calculating the root-mean-square radius of the
proton distribution relative to the center of mass, and thus the nuclear charge
radius (\textit{e.g.},
Refs.~\cite{bacca2012:6he-hyperspherical,caprio2014:cshalo}).

\subsection{Many-body space and symmetry structure}
\label{sec-background-space}

In symmetry-adapted formulations of the nuclear many-body
problem~\cite{dytrych2008:sp-ncsm}, based on the $\grpu{3}$ or $\grpsptr$
groups, intrinsic operators have a fundamental relationship to the structure of
the space in which many-body calculations are performed.  The very
definition of the symmetry-adapted many-body space and its truncation schemes
depend upon the existence of intrinsic realizations of the group generators, and
the calculational machinery may make use of the intrinsic group structure as
well.

The group $\grpsptr$ is generated by the full set of bilinear operators in
coordinates and/or momenta.  Elliott's $\grpu{3}$ subgroup, the harmonic
oscillator degeneracy group, is generated by those linear combinations of
bilinears which conserve the number of oscillator quanta.  This $\grpu{3}$ group decomposes as
$\grpu{3}=\grpu{1}\times\grpsu{3}$.  Here $\grpu{1}$ is simply the trivial
Abelian group of the harmonic oscillator Hamiltonian.  The generators of
$\grpsu{3}$ may be taken as the components $L_{1M}$ of the orbital angular
momentum and the components $\calQ_{2M}$ of a quadrupole tensor.  The orbital
angular momentum group $\grpso{3}$ is thus a subgroup.  The generators of
$\grpu{3}$ and $\grpsptr$ are reviewed in Sec.~\ref{sec-app-physical-ladder}.

In a symmetry-adapted approach, the basis for the nuclear many-body space is
chosen to reflect the organization of this space into irreducible
representations (irreps) either of the $\grpu{3}$ group chain
$\grpu{3}=\grpu{1}\times[\grpsu{3}\supset\grpso{3}]$ or the full $\grpsptr$
group chain $\grpsptr\supset\grpu{3}=\grpu{1}\times[\grpsu{3}\supset\grpso{3}]$.
A $\grpu{3}$ irrep is labeled by the $\grpu{1}$ quantum number $N$ together with
the $\grpsu{3}$ quantum numbers $(\lambda,\mu)$, which taken together form a
$\grpu{3}$ label $\omega=N(\lambda,\mu)$.  A $\grpu{3}$ irrep then decomposes
into $\grpso{3}$ irreps, labeled by the orbital angular momentum $L$.

The intrinsic generators for
$\grpu{3}$~\cite{kretzschmar1960:su3-shell-part2-com} and
$\grpsptr$~\cite{rosensteel1980:sp6r-shell} are obtained by the same
prescription described above, \textit{i.e.}, replacement of coordinates and
momenta by their intrinsic-frame expressions.
What is important is that, due to their bilinear structure, the laboratory-frame
generators $G$ decompose into the intrinsic operators $G'$ and center-of-mass
operators $\Gcm$, as
\begin{equation}
  \label{eqn-G-intrinsic-cm}
  G=G'+\Gcm.
\end{equation}
The two sets of operators $\lbrace G'\rbrace$ and  $\lbrace \Gcm \rbrace$ are
mutually commuting, and the operators within each set themselves respectively obey the
commutation relations to form the generators for intrinsic and center-of-mass
realizations of the group, that is, of $\grpu{3}$ or $\grpsptr$.

The implications of this intrinsic group structure are already familiar for the
$\grpso{3}$ orbital angular momentum subgroup.
Recall the classical result that the angular momentum about the center of mass and the angular
momentum of the center of mass add to give the total angular
momentum~\cite{landau1981:mechanics}.  In the quantum treatment, the existence of mutually commuting
intrinsic and center-of-mass angular momentum groups,
$\grpso[\text{intr}]{3}$ and $\grpso[\text{c.m.}]{3}$, combining to give the
total angular momentum group $\grpso{3}$ via the sum generators
$\Lvec=\Lvec'+\Lvec_{\text{c.m.}}$, ensures that the intrinsic and center-of-mass angular
momenta of the many-body system combine as $L'\times L_{\text{cm}}\rightarrow L$, according to the usual rules of angular
momentum addition, to yield the total angular momentum of
the system.

Similarly, the existence of an intrinsic $\grpu{1}$ group structure underlies
the traditional use of an $\Nmax$-truncated harmonic oscillator basis in
\textit{ab initio} no-core configuration interaction (NCCI)
calculations~\cite{barrett2013:ncsm}.  An $\Nmax$-truncated oscillator basis is
built from antisymmetrized products of harmonic oscillator single-particle
states and includes all product states with up to $\Nmax$ total harmonic oscillator
excitations relative to the lowest Pauli-allowed filling of oscillator
shells~\cite{mcgrory1975:spurious-com}.  With such a choice of basis, the
many-body space separates cleanly as the direct sum of a subspace which is free
of center-of-mass exitations and a complementary spurious space.  Within the
center-of-mass free subspace, all the wave functions factorize into
center-of-mass and intrinsic parts and share a well-defined motion in the
center-of-mass coordinate, described by a harmonic oscillator $0s$ ground
state wave function.  This zero-point motion in the center-of-mass coordinate
combines with some, in general, much more complicated structure in the intrinsic
coordinates, all while retaining fermionic antisymmetry.  (As a special case,
the $0\hw$ shell model space, consisting of the complete set of many-body states
defined in a single oscillator major shell, is well-known to have pure $0s$
center-of-mass motion~\cite{elliott1955:com-shell}.)

The $\Nmax$ truncation scheme works as it does as the result of the intrinsic
and center-of-mass harmonic oscillator $\grpu{1}$ groups,
$\grpu[\text{intr}]{1}$ and $\grpu[\text{c.m.}]{1}$, respectively, combining to
give the sum generator, or total number operator, $N=N'+\Ncm$ (see
Sec.~\ref{sec-app-ladder-intrinsic}).  Since the number operators $N'$, $\Ncm$,
and $N$ are mutually commuting, they may be simultaneously diagonalized in the
harmonic oscillator many-body basis, yielding (in principle) a new basis
consisting of simultaneous eigenstates of $N'$, $\Ncm$, and $N=N'+\Ncm$.  This
choice of basis decomposes the many-body space as a direct sum of subspaces with definite $N'$
and $\Ncm$, and truncation by $N$, as in the $\Nmax$ scheme, preserves such a
structure.

It is clearest to illustrate this transformation graphically, taking the example
of $\isotope[4]{He}$ with $\Nmax=4$.\footnote{For simplicity, we restrict
  attention to the even parity space, and thus even values of $N$, and
  furthermore restrict attention to intrinsic excitations of even parity, and
  thus even $N'$.  A parity-conserving Galilean-invariant Hamiltonian will
  not connect subspaces of even and odd $N'$.}  Each panel of
Fig.~\ref{fig:block-Nmax} represents the block structure of the matrix realization of
a Galilean-invariant intrinsic Hamiltonian operator, $H=T'+V$, as we restructure the basis,
which is indicated along the top edge of the matrix.
\begin{figure*}[t]
\begin{center}
\includegraphics[width=\ifproofpre{0.75}{0.75}\hsize]{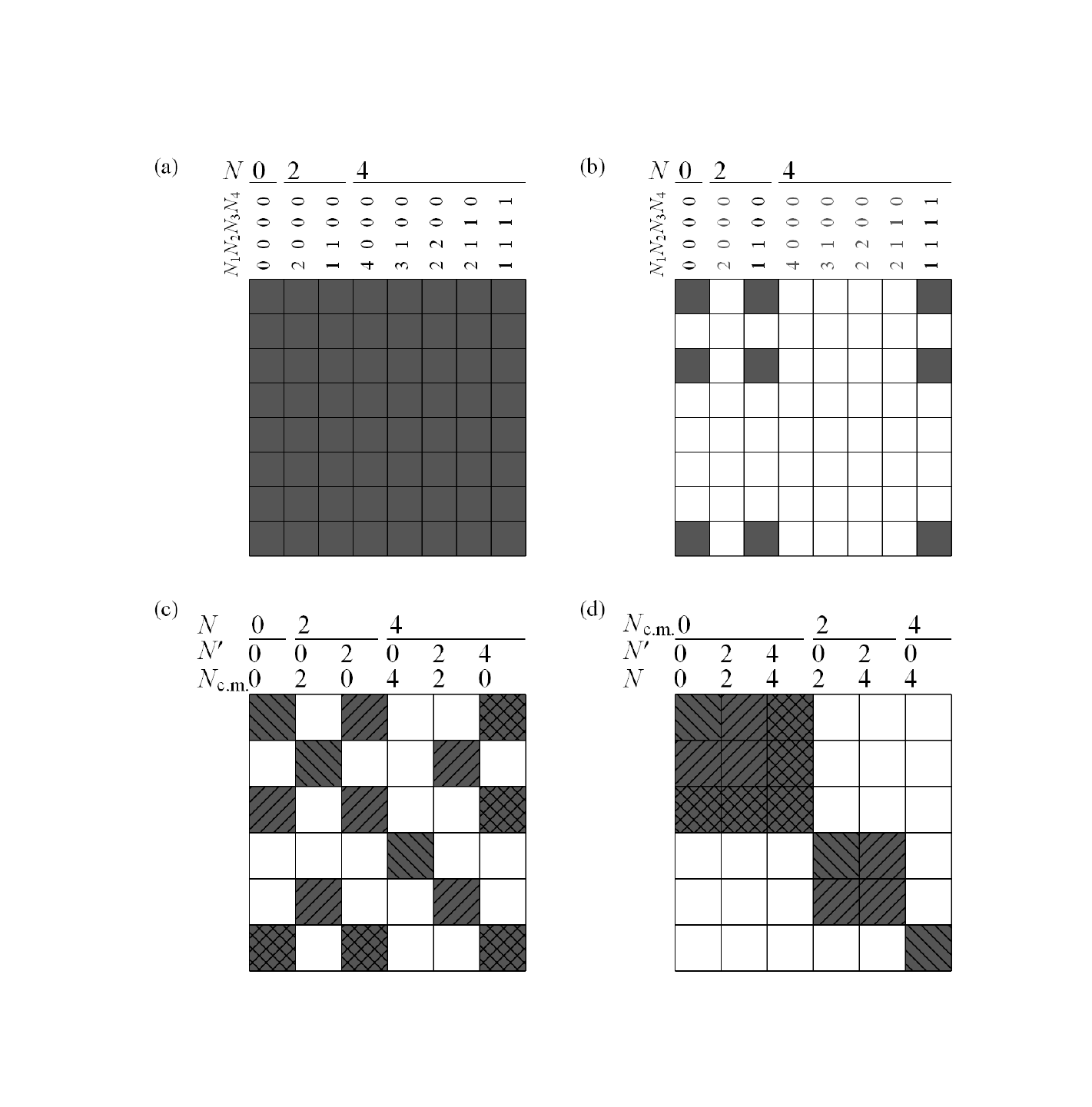}
\end{center}
\caption{Block structure of the matrix representation of a Galilean-invariant
  Hamiltonian, in the $\Nmax=4$ harmonic oscillator space of $\isotope[4]{He}$:
  (a)~Represented in the natural oscillator-configuration basis, with occupied
  oscillator shells $N_i$, grouped by $N$.  (b)~Similarly, but restricted to the
  FCI subspace with $N_i\leq1$.  (c)~Represented in a simultaneous eigenbasis of
  $N'$, $\Ncm$, and $N$, grouped by $N$.  (d)~Similarly, but grouped by $\Ncm$.
  Block sizes are not proportional to actual subspace dimensions (which are $1$ for $N=0$, $58$ for $N=2$, and $893$ for $N=4$ in an $M=0$ $M$-scheme~\cite{whitehead1977:shell-methods} basis for $\isotope[4]{He}$).
  Hatching indicates the intrinsic structure entering into the matrix element,
  namely, whether the matrix element involves bra and ket states with $N'=0$
  (backward hatched), $N'\leq2$ (forward hatched), or $N'\leq4$ (cross hatched).
\label{fig:block-Nmax}
}
\end{figure*}

The naturally constructed oscillator basis is obtained by distributing nucleons
over orbitals, each of which may be characterized by its major oscillator shell
$N_i$ (where, given indistinguishable particles, we may take $N_1\geq N_2 \geq
\cdots \geq N_A$ without loss of generality).  This basis is indicated in
Fig.~\ref{fig:block-Nmax}(a). The basis states are eigenstates of $N=\sum N_i$, and
are shown grouped by $N$ in Fig.~\ref{fig:block-Nmax}(a).  However, they are not eigenstates of $N'$ or $\Ncm$
separately, and there are thus no selection rules on a Galilean-invariant
operator.

Transforming to a basis obtained by simultaneously diagonalizing $N'$ and
$\Ncm$ yields the basis indicated in Fig.~\ref{fig:block-Nmax}(c).  These states are
obtained by a nontrivial unitary transformation of the original basis states
(\textit{i.e.}, taking linear combinations, not simply rearranging and
relabeling these states).  The basis states are still eigenstates of $N$, so the
unitary transformation takes place separately within each subspace of definite
$N$, and the basis states are again shown grouped by $N$ in
Fig.~\ref{fig:block-Nmax}(c).  A Galilean-invariant operator acts as the identity
operator on the center-of-mass degree of freedom and thus cannot connect states of
different $\Ncm$, \textit{i.e.}, $\tme{\Ncm'}{H'}{\Ncm}=0$ for $\Ncm'\neq\Ncm$.
This selection rule yields the block-sparse structure shown in
Fig.~\ref{fig:block-Nmax}(c).

Rearranging the basis to be grouped instead by $\Ncm$, as in
Fig.~\ref{fig:block-Nmax}(d), turns the block-sparse structure into a block-diagonal
structure.  This rearrangement of the basis defines a decomposition of the
$\Nmax=4$ space as a direct sum of subspaces, each the product of an
intrinsic subspace (successively more and more truncated) with a center-of-mass
subspace (involving successively higher $\Ncm$):
\begin{maybemultline}
\label{eqn:Nmax-decomposition}
\mathcal{H}^{(\Nmax=4)}
=
\lbrace0,2,4\rbrace_{\text{intr}}\times\lbrace0\rbrace_{\text{c.m.}}
\ifproofpre{\\}{}
+
\lbrace0,2\rbrace_{\text{intr}}\times\lbrace2\rbrace_{\text{c.m.}}
+
\lbrace0\rbrace_{\text{intr}}\times\lbrace4\rbrace_{\text{c.m.}}
.
\end{maybemultline}
The Galilean invariance of the operator dictates that matrix elements between
states with identical intrinsic structure are identical, regardless of the
spectator center-of-mass wave function.  Therefore, each successive block along
the diagonal effectively represents a truncation, or submatrix, of the
preceeding block.\footnote{Details of the actual choice of
  basis states within each subspace, in particular, imposing angular momentum
  coupling between the intrinsic and center-of-mass motion, could obscure this
  simple relationship.}  This structure is emphasized by the hatch lines, in
Fig.~\ref{fig:block-Nmax}(d), which indicate the intrinsic quantum numbers entering
into each block of the matrix.

The desired center-of-mass free states (with center-of-mass $0s$ wave functions)
arise from diagonalization of the first block along the diagonal.  The remaining
blocks yield the spurious states, which also reflect a poorer description of the
intrinsic structure, obtained in a more severely truncated subspace.
For instance, considering the decomposition of the $\isotope[4]{He}$
$\Nmax=4$ space in~(\ref{eqn:Nmax-decomposition}), the spectrum of $H'$ in this
space consists of three copies of the intrinsic excitation spectrum, each a
truncated approximation to the untruncated intrinsic spectrum: the
highest-fidelity rendition of the intrinsic spectrum, obtained in the
$\lbrace0,2,4\rbrace_{\text{intr}}$ space and accompanied by no center-of-mass
excitation; an intermediate-fidelity rendition of the intrinsic spectrum,
obtained in the $\lbrace0,2\rbrace_{\text{intr}}$ space and accompanied by
$\Ncm=2$ excitations; and a lowest-fidelity intrinsic spectrum, obtained in the
$\lbrace0\rbrace_{\text{intr}}$ space and accompanied by $\Ncm=4$ excitations.

It is to be emphasized that, in practical calculations, the basis states of
definite $N'$ and $\Ncm$ [Fig.~\ref{fig:block-Nmax}(c,d)] are never explicitly
constructed.  Rather, the mere fact that these basis states exist underlies the
decomposition~(\ref{eqn:Nmax-decomposition}) of the truncated many-body space
and thus the emergence of factorized eigenstates of definite $\Ncm$.  These
factorized eigenstates simply come out of the diagonalization of the
Galilean-invariant Hamiltonian.

In particular, to reap the benefits of the structure of the space, the
Hamiltonian must be chosen so as to respect the block structure shown in
Fig.~\ref{fig:block-Nmax}(d).  The original one-body kinetic energy $T$ connects
oscillator basis states involving different center-of-mass excitations,
\textit{i.e.}, $\tme{\Ncm'}{T}{\Ncm}\neq0$, and in fact yields a
block-tridiagonal structure in $\Ncm$.  It is this consideration that dictates
its replacement by $T'$, that is, use of an intrinsic Hamiltonian $H'$, in NCCI
calculations.  The resulting low-lying spectrum then reflects the block
structure described above, yielding multiple copies of the intrinsic spectrum,
and these may be expected to appear at low energy, as the center-of-mass
excitations do not carry any intrinsic kinetic energy.  The
spurious states carry no useful additional information about the intrinsic
structure, serving only to pollute the calculated spectrum.  However, the center-of-mass
excited states may still be shifted out of the low-lying spectrum by addition of
a center-of-mass Lawson
term~\cite{gloeckner1974:spurious-com,whitehead1977:shell-methods,lawson1980:shell},
proportional to $\Ncm$, which preserves the block structure shown in
Fig.~\ref{fig:block-Nmax}(d).  See Fig.~8 of Ref.~\cite{caprio2012:csbasis} for
an illustration both of the structure of the spurious spectrum and of the effect
of the Lawson term.

In contrast, in calculations which depart from an $\Nmax$-truncated oscillator
many-body space, either by generalizing the oscillator truncation
scheme~\cite{vary2018:gentrunc-ostuka17} or starting from non-oscillator
single-particle
orbitals~\cite{caprio2012:csbasis,constantinou2017:natorb-natowitz16}, the
decomposition~(\ref{eqn:Nmax-decomposition}) of the space in general may be
expected to break down.  This breakdown is illustrated for a full
configuration-interaction (FCI) basis (\textit{e.g.},
Ref.~\cite{abe2012:fci-mcsm-ncfc}) in Fig.~\ref{fig:block-Nmax}(b), where harmonic
oscillator configurations are taken subject to a single-particle cutoff $N_i\leq
1$ on the occupied oscillator shells.  The resulting space is a subspace of the
$\Nmax=4$ space.  However, we no longer have a complete basis for each subspace
of fixed $N$ (in particular, for the $N=2$ and~$4$ subspaces), which precludes
the unitary transformation to the basis of Fig.~\ref{fig:block-Nmax}(c).

Nonetheless, even without the exact factorization ensured by an
$\Nmax$-truncated oscillator basis, an approximate factorization of $s$-wave
center-of-mass motion in the calculated eigenstates can still be
obtained~\cite{hagen2009:coupled-cluster-com,*hagen2010:coupled-cluster,caprio2012:csbasis,constantinou2017:diss}.
Even though the center-of-mass wave function is not directly accessible in an
antisymmetrized products basis, to the extent that the center-of-mass wave
function resembles $0s$ harmonic oscillator zero-point motion, for some choice
of the oscillator length parameter $\bcm$ (Sec.~\ref{sec-app-ladder-intrinsic}),
this situation may be recognized by evaluating
$\tbracket{\Ncm}$~\cite{hagen2009:coupled-cluster-com,*hagen2010:coupled-cluster,caprio2012:csbasis,constantinou2017:diss}.
Approximate center-of-mass factorization may arise spontaneously, or it may be
coerced by addition of a Lawson term, but doing so generally comes at the
expense of convergence of the intrinsic wave function (fidelity to $0s$
center-of-mass motion must be traded off against fidelity of the intrinsic wave
function).  See, \textit{e.g.}, Fig.~9 of Ref.~\cite{caprio2012:csbasis}
for an illustration of the resulting spectrum and use of the Lawson term for
calculations in an antisymmetrized product basis of Laguerre
functions~\cite{shull1955-continuum,weniger1985:fourier-plane-wave,mccoy2016:lgalg}.

With this understanding of the intrinsic oscillator structure of the many-body
space in hand, let us now proceed to $\grpu{3}$.  The intrinsic group
$\grpu[\text{intr}]{3}$ and center-of-mass group $\grpu[\text{c.m.}]{3}$ are
likewise mutually commuting.  Consequently, irreps $\omega'$ and
$\omegacm$, describing the intrinsic and center-of-mass motion,
respectively, combine as
$\omega'\times\omegacm\rightarrow\omega$ according to
the usual $\grpu{3}$ coupling rules (\textit{e.g.},
Ref.~\cite{wybourne1974:groups}).  These rules reduce to the addition of the
$\grpu{1}$ quantum numbers [$N=N'+\Ncm$] as above, and combination of the
$(\lambda,\mu)$ quantum numbers according to the $\grpsu{3}$ coupling
rules~\cite{oreilly1982:su3-coupling}.

The existence of mutually commuting intrinsic and center-of-mass $\grpu{3}$
groups ensures that, if the many-body space is truncated according to $\grpu{3}$
quantum numbers, a decomposition of the space into products of intrinsic and
center-of-mass subspaces can again be carried out.  The many-body space can be
simultaneously decomposed into irreps of the intrinsic, center-of-mass, and
total $\grpu{3}$ groups (this may be thought of as simultaneously diagonalizing
the Casimir operators for the three groups), and a basis can (in principle) be
obtained consisting of states of definite $\omega'$, $\omegacm$, and $\omega$.
The decomposition in Fig.~\ref{fig:block-Nmax}(c) is then broken down more finely by
inclusion of additional quantum numbers, but the reorganization into
block-diagonal form by $\Ncm$, as in Fig.~\ref{fig:block-Nmax}(d), can still be
carried out.  Thus, notably, the factorization of the resulting wave functions
into intrinsic and center-of-mass factors, and the extraction of center-of-mass
free states, can again be obtained, even if the space is truncated not just by
total $N$ but more finely by total
$\omega$~\cite{kretzschmar1960:su3-shell-part2-com,verhaar1960:shell-com,hecht1971:su3-com,millener1975:14b-beta-su3,millener1992:su3-multi-shell,luo2013:su3cmf}.

In particular, the natural basis for a $\grpu{3}$-coupled NCCI calculation, as in the
symmetry-adapted NCSM (SA-NCSM) of
Refs.~\cite{dytrych2013:su3ncsm,dytrych2016:su3ncsm-12c-efficacy}, is obtained
by distributing nucleons over oscillator shells, yielding definite total $N$,
then ensuring that the resulting states are coupled to give good total $\omega$.
The separability property guarantees that factorized center-of-mass motion will
be obtained if an intrinsic Hamiltonian (with optional Lawson term) is
diagonalized in any such basis selected to include all basis states of a given
set of total $\grpu{3}$ quantum numbers $\omega$.  Diagonalization in the traditional
$\Nmax$-truncated oscillator space is recovered as the special case in which all
$\grpu{3}$ irreps up to a given $N$ are retained.
\begin{figure*}[t]
\begin{center}
\includegraphics[width=\ifproofpre{0.75}{0.75}\hsize]{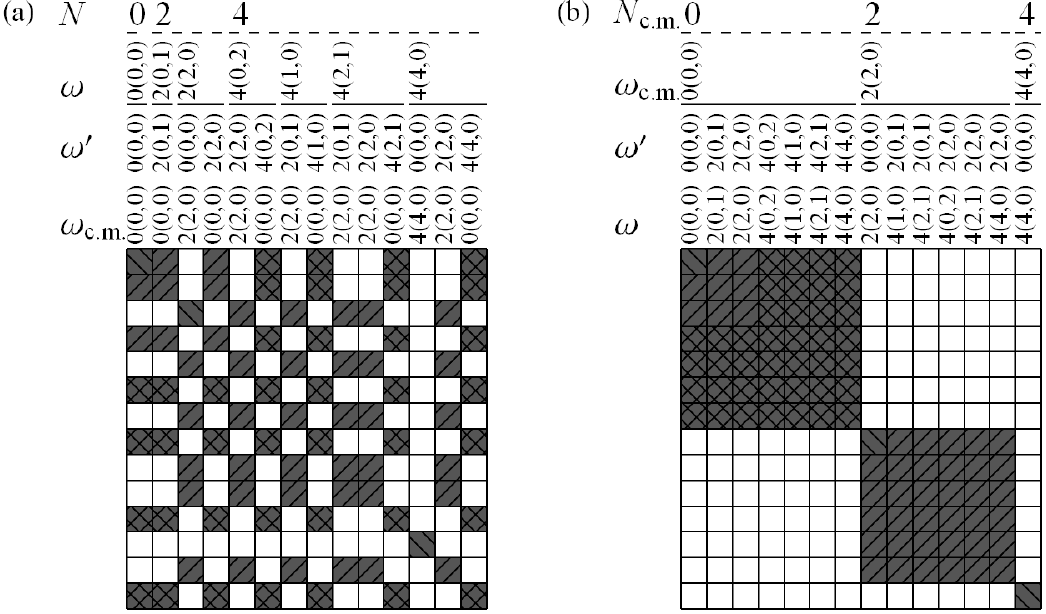}
\end{center}
\caption{Block structure of the matrix representation of Galilean-invariant
  Hamiltonian, in the $\Nmax=4$ harmonic oscillator space of $\isotope[4]{He}$, further decomposed into $\grpu{3}$ subspaces:
  (a)~Represented in a basis of definite $\omega'$, $\omegacm$ (or, equivalently, $\Ncm$), and $\omega$, grouped by $\omega$.
  (b)~Similarly, but grouped by $\omegacm$ (or, equivalently, $\Ncm$).
  Hatching indicates whether the matrix element involves bra and ket states with $N'=0$
  (downward sloped hatching), $N'\leq2$ (upward sloped hatching), or $N'\leq4$ (cross hatching).
\label{fig:block-Nmax-u3}
}
\end{figure*}

We illustrate the reorganization of the many-body space by intrinsic and total
$\grpu{3}$ quantum numbers, again for the $\Nmax=4$ space of $\isotope[4]{He}$,
in Fig.~\ref{fig:block-Nmax-u3}.  For the intrinsic $\grpu{3}$, the possible
irreps with $N'\leq4$ (even) have $\omega'=0(0,0)$, $2(0,1)$, $2(2,0)$,
$4(0,2)$, $4(1,0)$, $4(2,1)$, and $4(4,0)$, and, as it turns out, the same
values are obtained for the $\omega$ of the total $\grpu{3}$.\footnote{These
  values may be read off from Fig.~5(a) of Ref.~\cite{luo2013:su3cmf}.}  For the
center-of-mass degree of freedom, the $\grpu{3}$ quantum numbers are trivially
related to the number of oscillator quanta, as $\omegacm=\Ncm(\Ncm,0)$.  The
only nontrivial $\grpu{3}$ couplings entering into Fig.~\ref{fig:block-Nmax-u3}
are for the coupling of the $N'=2$ intrinsic irreps to the $\Ncm=2$
center-of-mass motion: $2(0,1)\times2(2,0)\rightarrow4(1,0),4(2,1)$ and
$2(2,0)\times2(2,0)\rightarrow4(0,2),4(2,1),4(4,0)$.  The resulting basis states
are grouped by $\omega$ in Fig.~\ref{fig:block-Nmax-u3}(a), giving block-sparse
structure for a Galilean-invariant operator.  These basis states are then
rearranged by $\Ncm$ (or, equivalently, $\omegacm$) in
Fig.~\ref{fig:block-Nmax-u3}(b), giving block-diagonal structure for a
Galilean-invariant operator.  Observe that truncating the $\Nmax=4$ space by
$\omega$, for instance, eliminating the $\omega=4(0,2)$ and~$4(1,0)$ $\grpu{3}$
subspaces, would reduce (subset) the set of $\omega$ values included in
Fig.~\ref{fig:block-Nmax-u3}(a), and would thus eliminate certain rows and
columns from within the block-diagonal structure of
Fig.~\ref{fig:block-Nmax-u3}(b).  However, it would not destroy this
block-diagonal structure, nor would it interfere with the decomposition of the
space into intrinsic and center-of-mass product subspaces that this structure
implies.

In the SpNCCI
framework~\cite{mccoy2018:diss,mccoy2018:spncci-busteni17,mccoyxxxx:spfamilies},
the nuclear many-body calculation is carried out in a center-of-mass free
$\grpsptr$ basis.  Matrix elements of the intrinsic Hamiltonian (and other
intrinsic operators for observables) in this basis are computed using a
recurrence relation derived from the commutation relations between the intrinsic
$\grpsptr$ generators and certain $\grpsu{3}$-coupled unit tensor operators
which provide a basis for the space of operators.  The approach builds on ideas
of Reske, Suzuki, and
Hecht~\cite{reske1984:diss,suzuki1986:sp6r-alpha-cluster-me,suzuki1986:sp6r-cluster}.

Seed matrix elements for the recurrence are obtained via the usual
second-quantized approach, on $\grpu{3}$-coupled SA-NCSM basis states with
definite $0s$ center-of-mass motion (these may be obtained either by
diagonalizing the center-of-mass number operator $\Ncm$ or solving for the null
space of the center-of-mass annihilation operator $\cveccm$, defined in
Sec.~\ref{sec-app-ladder-intrinsic}).  The recurrence process then bypasses the
need for any further reference to the laboratory-frame single-particle
representation of the many-body problem.  Since the seed matrix elements involve
center-of-mass free states, and since the recurrence is obtained by implicitly
acting on these states with intrinsic $\grpsptr$ raising generators
$A^{\prime\,(2,0)}$, the resulting matrix elements are those on a
  center-of-mass free $\grpsptr$ basis, although this basis need never be explicitly
  constructed in terms of laboratory-frame states. The calculations rely upon the
  relations, developed in the present work, between the defining expressions for the intrinsic
  bilinear operators and their explicit two-body forms~\cite{mccoy2018:diss}.

Regardless of how ``center-of-mass free'' solutions to the many-body problem are
obtained, whether as a byproduct of diagonalizing a Galilean-invariant
Hamiltonian or \textit{a priori} through construction of a center-of-mass free
basis for the problem, it is important to note that imposing a well-defined $0s$
zero-point center-of-mass motion is not to be conflated with complete removal of
the center-of-mass degree of freedom from the problem.  Unless care is taken to
work in terms of intrinsic operators for observables, calculated observables may
in general be expected to reflect contamination from this zero-point
center-of-mass motion.  A case in point is provided by the r.m.s.\ radius, as
discussed in Sec.~\ref{sec-background-obs}.  The squared radius evaluated with
the laboratory-frame operator, in a center-of-mass free state, measures a
density which is the convolution of the density in the intrinsic frame with the
Gaussian profile of the zero-point motion of the center of
mass~\cite{cockrell2012:li-ncfc}.  In this case, the required correction is
straightforward [see~(\ref{eqn-rsqr-decomposition})].

While there are certain restricted circumstances,
involving center-of-mass free wave functions, under which the naive one-body
laboratory-frame transition operators can be used (as they indeed often are)
without introducing center-of-mass contamination to calculated observables,
even here, an understanding these circumstances requires an understanding of the
intrinsic forms of these operators (see Secs.~\ref{sec-dipole}
and~\ref{sec-bilinear}).



\section{Definitions: Operators and coordinates}
\label{sec-formalism}

\subsection{One-body and two-body operators}
\label{sec-nbody}

One-body operators are operators on the $A$-particle space, but they are uniquely
defined by their action on the single-particle space. Similarly, two-body
operators are operators on the $A$-particle space, but they are uniquely defined by their
action on the two-particle space.  For fermionic problems, in particular, the
matrix elements of one-body or two-body operators in a basis of antisymmetrized
product states (Slater determinants) on the $A$-particle space can readily be
computed from the matrix elements of these operators between basis states for
the one-particle or two-particle spaces, respectively, through second
quantization (see, \textit{e.g.}, Ref.~\cite{negele1988:many-particle}).  Thus,
in setting up a many-body calculation involving these operators, it is only
necessary to evaluate the relevant matrix elements on the one-particle or
two-particle spaces, respectively.  These one-body and two-body matrix elements
then serve as the input to the standard computational machinery of the many-body
calculation, which can generate the matrix elements of these operators in the
$A$-particle antisymmetrized product basis.

As our focus in this work is to establish relations between different one-body
and two-body decompositions of intrinsic operators, we first, in this section,
establish definitions for these operators.  In particular, we must have an
unambiguous notation for relating operators on the one-particle and two-particle spaces
with the corresponding one-body and two-body operators on the $A$-particle space,
respectively.
A \textit{one-body operator}, acting on the $A$-particle space,
is obtained by taking an operator $u$ on the single-particle space and applying
it uniformly to all $A$ particles:
\begin{equation}
\label{eqn-obo}
U[u]=\sum_i u_i,
\end{equation}
where $u_i$ acts as $u$ on the single-particle space of the $i$th particle and
as the identity for all other particles.

Similarly, a \textit{two-body operator}, acting on the $A$-particle space,
is obtained by taking an operator $v$ on the two-particle space and applying
it uniformly to all pairs of $A$ particles:
\begin{equation}
\label{eqn-tbo}
V[v]=\tfrac12 \sumprime_{ij} v_{ij},
\end{equation}
where $v_{ij}$ acts as $v$ on the two-particle space of the $i$th and $j$th particles and
as the identity for all other particles.
Here the prime on the sum indicates omission of diagonal terms (\textit{i.e.},
summation over $i,j=1,\ldots,A$ subject to the restriction $i\neq j$).  In this
definition, the condition is imposed that the operator $v$ must be symmetric
under interchange of particles in the two-particle space ($v_{12}=v_{21}$), to
ensure that $V[v]$ is symmetric under interchange of particle indices in the
$A$-particle space.

While the distinction between one-body and two-body operators is clear when the
operators are acting on the full Fock space, with arbitrary number of particles,
the distinction is not as clear on a space of fixed particle number $A$.  In
fact, any one-body operator may be ``upgraded'' to an equivalent two-body
operator via the summation identities in Appendix~\ref{sec-app-sum}.  In
particular, the one body operator $U[u]$ may be reexpressed
[see~(\ref{eqn-upgrade})] as\footnote{This relation~(\ref{eqn-upgrade-U})
  between $U[u]$ and $V[u_1+u_2]$ is an $A$-dependent relation, valid separately
  on each $A$-particle space, not a true $A$-independent operator identification.
  Thus, \textit{e.g.}, if $U[u]$ is to be treated as a two-body operator in a
  many-body computation, the two-body matrix elements evaluated for the $A=2$
  basis must then be rescaled by $1/(A-1)$ to be applied on the $A$-particle
  problem.  }
\begin{equation}
\label{eqn-upgrade-U}
U[u]=\frac{1}{(A-1)}V[u_1+u_2],
\end{equation}
on the $A$-particle space, where $u_1+u_2\,(= v_{12})$ is the operator on the
two-particle space obtained by the one-body action of $u$ on this space.  (However,
the converse does not hold, in that a two-body operator cannot, in general, be
reduced to an equivalent one-body operator.)  Such interconversion between
one-body and two-body forms of an operator are essential to the relations
obtained in this work.


\subsection{Relative coordinates (two-body system)}
\label{sec-relative}

Relative and c.m.\ coordinates for the two-body system are proportional to the
difference $\xvec_1-\xvec_2$ and sum $\xvec_1+\xvec_2$ of single-particle
coordinates, respectively, and similarly for the momenta.  However, there is
freedom in the choice of conventional factors in these definitions, arising from
the freedom to carry out a canonical transformation on the coordinates and
momenta.  We therefore review the definitions so that we may avoid ambiguity in
the transformation between a relative operator on the two-particle space (which will
be expressed in terms of the relative coordinate $\xvecrel$ and momentum
$\kvecrel$) and the corresponding intrinsic operator on the $A$-particle space
obtained via~(\ref{eqn-tbo}), as in the example~(\ref{eqn-tbo-ksqr}).

In traditional mechanics applications, it is natural to take the relative
coordinate $\xvecrel$ as the displacement between particles and the
c.m.\ coordinate $\xveccm$ as the mean of the coordinate vectors, that is, quite
literally the ``center of the masses'' (at least in the case of equal masses,
which we consider here):
\begin{equation}
\label{eqn-coords-rcm-mechanics}
\begin{aligned}
\xvecrel&=\phantom{\tfrac12}(\xvec_1-\xvec_2)& \quad \xveccm&=        {\tfrac12}(\xvec_1+\xvec_2)\\
\kvecrel&=        {\tfrac12}(\kvec_1-\kvec_2)& \quad \kveccm&=\phantom{\tfrac12}(\kvec_1+\kvec_2).
\end{aligned}
\end{equation}
The momentum $\kveccm$ conjugate to the c.m.\ coordinate is then simply the total
momentum of the system (the normalizations for the conjugate momenta, $\kvecrel$
and $\kveccm$, are forced from the definitions of $\xvecrel$ and $\xveccm$, by
the requirement of conjugacy).

However, in working with the quantum harmonic oscillator, it is natural to
recognize the duality between coordinates and momenta more explicitly, by
symmetrically distributing the coefficients of $1/2$ between the coordinate and
the momentum, as adopted in, \textit{e.g.},
Ref.~\cite{moshinsky1996:oscillator}.  This also provides greater parallelism
between the treatment of the relative and c.m.\ degrees of freedom:
\begin{equation}
\label{eqn-coords-rcm-symmetric}
\begin{aligned}
\xvecrel&={\tfrac1{\sqrt2}}(\xvec_1-\xvec_2)& \quad \xveccm&={\tfrac1{\sqrt2}}(\xvec_1+\xvec_2)\\
\kvecrel&={\tfrac1{\sqrt2}}(\kvec_1-\kvec_2)& \quad \kveccm&={\tfrac1{\sqrt2}}(\kvec_1+\kvec_2).
\end{aligned}
\end{equation}

For results which are dependent upon the choice of
convention~(\ref{eqn-coords-rcm-mechanics}) or~(\ref{eqn-coords-rcm-symmetric})
for the relative-c.m.\ coordinates and momenta, we shall indicate the appropriate
coefficients obtained under each of the two conventions in braces, with the upper value
for mechanics convention and the lower value for the symmetric convention.  Thus,
\textit{e.g.}, the definitions~(\ref{eqn-coords-rcm-mechanics}) or~(\ref{eqn-coords-rcm-symmetric}) may be expressed together as
\begin{equation}
\label{eqn-coords-rcm}
\begin{aligned}
\xvecrel&=\Bstack{       1}{\tfrac1{\sqrt2}}(\xvec_1-\xvec_2)& \quad  \xveccm&=\Bstack{\tfrac12}{\tfrac1{\sqrt2}}(\xvec_1+\xvec_2)\\
\kvecrel&=\Bstack{\tfrac12}{\tfrac1{\sqrt2}}(\kvec_1-\kvec_2)& \quad  \kveccm&=\Bstack{       1}{\tfrac1{\sqrt2}}(\kvec_1+\kvec_2).
\end{aligned}
\end{equation}

Given that we will be focusing on Galilean-invariant systems and intrinsic
observables, we will have special interest in \textit{relative} two-body
operators.  These are two-body operators defined in terms of an operator $v$ on
the two-particle space which is itself Galilean-invariant and, thus, if represented
in terms of relative and center-of-mass (c.m.) coordinates
(Sec.~\ref{sec-relative}), can only involve the relative coordinate, not the
c.m.\ coordinate.  As a concrete illustration of the notation~(\ref{eqn-tbo})
for two-body operators, consider the squared relative momentum operator on the
two-body system, which is, to within the conventional factors
defined above in~(\ref{eqn-coords-rcm}), given by
$\kvecrel^2\propto(\kvec_1-\kvec_2)^2$.  Then the corresponding two-body operator obtained on the
$A$-particle space is
\begin{equation}
\label{eqn-tbo-ksqr}
V[\kvecrel^2]\propto V[(\kvec_1-\kvec_2)^2]=\tfrac12\sumprime_{ij}(\kvec_i-\kvec_j)^2.
\end{equation}


\subsection{Intrinsic coordinates (many-body system)}
\label{sec-intrinsic}

We now turn to coordinates for the $A$-particle system.  These are, naturally,
denoted by $\xvec_i$ ($i=1,\ldots,A$), with conjugate momenta $\vec{p}_i$
($i=1,\ldots,A$) or, as we shall equivalently use in their place in the
following discussion, wave vectors $\kvec_i$ ($i=1,\ldots,A$), where
$\vec{p}_i=\hbar\kvec_i$.  However, to define an observable in terms of
coordinates and momenta, in such a way that it is manifestly Galilean-invariant
quantity, we work instead with \textit{intrinsic coordinates} $\xvec'_i$ and
\textit{intrinsic momenta} $\kvec'_i$ defined with respect to the system's
center-of-mass frame.

The defining property of these intrinsic coordinates and momenta is that
$M^{-1}\sum_i m_i\xvec'_i=0$ (\textit{i.e.}, the center of mass is at the
origin) and $\sum_i \kvec'_i=0$ (\textit{i.e.}, the total momentum of the system
vanishes), where $m_i$ are the masses, and $M=\sum_i m_i$.  The transformation
to intrinsic coordinates and momenta is obtained, for general choices of the
masses $m_i$, as~\cite{landau1981:mechanics}
\begin{equation}
\label{eqn-coords-intrinsic-mass}
\begin{aligned}
\xvec'_i&=\xvec_i-\phantom{\frac{m_i}{M}}\xveccm         & \quad \xveccm&=         {\frac{1}{M}} \sum_i m_i \xvec_i \\
\kvec'_i&=\kvec_i-        {\frac{m_i}{M}}\kveccm         & \quad \kveccm&= \phantom{\tfrac{1}{M}} \sum_i \kvec_i.
\end{aligned}
\end{equation}
However, for simplicity, we specialize to particles of equal mass, namely,
taking $m_i\rightarrow m_N$ and $M\rightarrow Am_N$ (see footnote
\ref{footnote-pn-mass}).  We then have,
\begin{equation}
\label{eqn-coords-intrinsic-mechanics}
\begin{aligned}
\xvec'_i&=\xvec_i-\phantom{\frac{1}{A}}\xveccm         & \quad \xveccm&=         {\frac{1}{A}} \sum_i \xvec_i \\
\kvec'_i&=\kvec_i-        {\frac{1}{A}}\kveccm         & \quad \kveccm&= \phantom{\frac{1}{A}} \sum_i \kvec_i.
\end{aligned}
\end{equation}

Note that $\xveccm$ and $\kveccm$ may alternatively be defined, by canonical
transformation, such that the factor of $1/A$ is symmetrically distributed
between them (see also Sec.~\ref{sec-relative}), giving
\begin{equation}
\label{eqn-coords-intrinsic-symmetric}
\begin{aligned}
\xvec'_i&=\xvec_i-        {\frac{1}{\sqrt A}}\xveccm         & \quad \xveccm&=         {\frac{1}{\sqrt A}} \sum_i \xvec_i \\
\kvec'_i&=\kvec_i-        {\frac{1}{\sqrt A}}\kveccm         & \quad \kveccm&=         {\frac{1}{\sqrt A}} \sum_i \kvec_i.
\end{aligned}
\end{equation}
For results which are dependent upon the choice of
convention~(\ref{eqn-coords-intrinsic-mechanics})
or~(\ref{eqn-coords-intrinsic-symmetric}) for the c.m.\ coordinates and momenta,
we shall indicate the appropriate coefficients obtained under each of the two
conventions in braces, with the upper value for mechanics convention and the
lower value for the symmetric convention.  Thus, \textit{e.g.}, the
definitions~(\ref{eqn-coords-intrinsic-mechanics})
or~(\ref{eqn-coords-intrinsic-symmetric}) themselves may be expressed together
as
\begin{equation}
\label{eqn-coords-intrinsic}
\begin{aligned}
\xvec'_i&=\xvec_i- \Bstack{           1}{\sfrac{1}{\sqrt{A}}} \xveccm         & \quad \xveccm&= \Bstack{\sfrac{1}{A}}{\sfrac{1}{\sqrt{A}}} \sum_i \xvec_i \\
\kvec'_i&=\kvec_i- \Bstack{\sfrac{1}{A}}{\sfrac{1}{\sqrt{A}}} \kveccm         & \quad \kveccm&= \Bstack{           1}{\sfrac{1}{\sqrt{A}}} \sum_i \kvec_i.
\end{aligned}
\end{equation}
However, either way, eliminating the intermediate reference to the c.m.\ coordinate and momentum gives simply
\begin{equation}
\label{eqn-coords-intrinsic-explicit-sum}
\begin{aligned}
\xvec'_i&=\xvec_i-        {\frac{1}{A}}  \sum_j \xvec_j \\
\kvec'_i&=\kvec_i-        {\frac{1}{A}}  \sum_j \kvec_j
\end{aligned}
\end{equation}
for the intrinsic coordinates and momenta.

Note that the intrinsic coordinates $\xvec'_i$ and momenta $\kvec'_i$ do
\textit{not} satisfy canonical commutation relations (see
Sec.~\ref{sec-app-ladder-intrinsic}).  Also, although the intrinsic coordinates
$\xvec'_1$ or $\xvec'_2$ on the two-particle space are simply related to the
relative coordinate $\xvecrel$, one should not mistakenly presume that intrinsic
coordinates simply reduce to relative coordinates on the two-particle space.


\section{Dipole operator}
\label{sec-dipole}

It is instructive to first consider the dipole or $E1$
operator~\cite{eisenberg1988:v2}, as a simpler case, before moving on to the
bilinear operators which serve as the main focus of the present work.  The
dipole operator allows us to illustrate several ingredients which arise in the
treatment of bilinear operators, but without the distraction of some more
cumbersome algebra.  In particular, the following discussion of the dipole
operator provides examples of transforming an operator to intrinsic coordinates,
decomposing it into isoscalar and isovector contributions, decomposing an
intrinsic operator into the
``naive'' one-body operator and a c.m.\ ``recoil'' contribution, and
expressing an intrinsic operator as a pure relative two-body operator.

The mass, or isoscalar, dipole operator, which we define as the one-body
operator
\begin{equation}
\label{eqn-D-obo}
\Dvec=U[\xvec]=\sum_i \xvec_i,
\end{equation}
is simply proportional to the c.m.\ coordinate $\xveccm$~(\ref{eqn-coords-intrinsic}) of the $A$-body system.
It therefore comes as no surprise that, in the intrinsic frame, this operator
vanishes identically.  That is, if we attempt to construct an intrinsic dipole operator $\Dvec'$
by the substitution $\xvec_i\rightarrow\xvec_i'$~(\ref{eqn-coords-intrinsic}), we obtain
\begin{equation}
\label{eqn-D-intr}
\begin{aligned}
\Dvec'&=\sum_i \bigl(\xvec_i-\frac{1}{A}\sum_j\xvec_j\bigr)\\
&=\sum_i \xvec_i-\frac{1}{A}\bigl(\sum_i1)\bigl(\sum_j\xvec_j)\\
&=0,
\end{aligned}
\end{equation}
where we have recognized that summing over a ``free'' particle index which
does not appear in the summand simply
introduces a counting factor $A$ ($\sum_i1=A$).

Let us therefore move on to the variants of the dipole operator which
distinguish protons and neutrons.  We may define a proton dipole operator
$\Dvec_p$, in which the summation over particles runs only over protons, as
$\Dvec_p=\sum_{i\in p} \xvec_i$ (this proton dipole operator is, to within
multiplication by the electric charge $e$, the physical electric dipole
operator).  We may similarly define a neutron dipole operator $\Dvec_n$, in
which the summation runs only over neutrons, as $\Dvec_n=\sum_{i\in n} \xvec_i$.
However, by invoking a
restricted sum over particles, we have violated the defining property~(\ref{eqn-obo}) of a one-body
operator, as one which takes the action of an
operator defined on the single-particle space, and sums this action over
\textit{all} $A$ particles. We thus instead
allow the summation to range over all $A$ particles, but modify
the definition of the operator on the single-particle space, so that it ``sees''
only protons, and vanishes when acting on neutrons, or \textit{vice versa}.
That is, we have
\begin{equation}
\label{eqn-D-pn-obo}
\begin{aligned}
\Dvec_p&=U[\xvec\delta_p]&&=\sum_i \xvec_i\delta_{i,p}\\
\Dvec_n&=U[\xvec\delta_n]&&=\sum_i \xvec_i\delta_{i,n},
\end{aligned}
\end{equation}
where $\delta_p$ is an operator defined on the single-particle space, which acts
as the identity on a proton and vanishes acting on a neutron, and \textit{vice
  versa} for $\delta_n$.  Therefore, in the sum, $\delta_{i,p}=1$ if the $i$th
particle is a proton or $0$ if the particle is a neutron, and \textit{vice
  versa} for $\delta_{i,n}$.

The isospin formalism provides the natural framework for defining such proton
and neutron selection operators, and then allows us to decompose the proton and
neutron dipole operators into isoscalar and isovector contributions.  Following
common practice in nuclear theory, we work with the Pauli matrix operators
$\vec{\tau}$ for the nucleons; these are twice the isospin operators
($\vec{\tau}=2\vec{t}$).  We adopt the convention that $t_3=+1/2$ for the proton
and $t_3=-1/2$ for the neutron, so $\tau_0\,(=\tau_3)$ has eigenvalues $+1$ for
the proton and $-1$ for the neutron.  Then
$\delta_p=\tfrac12(1+\tau_0)$ and $\delta_n=\tfrac12(1-\tau_0)$.
The expressions for proton and neutron dipole operators
in~(\ref{eqn-D-pn-obo}) consolidate to
\begin{equation}
\label{eqn-D-alpha-obo}
\Dvec_\alpha=U[\xvec\delta_\alpha]=\sum_i \xvec_i\delta_{i,\alpha},
\end{equation}
where we combine these definitions as
\begin{equation}
\label{eqn-delta-alpha}
\delta_\alpha=\tfrac12(1+\alpha\tau_0),
\end{equation}
and use $\alpha=+1$ to select protons or $\alpha=-1$ to select neutrons.

This proton or neutron dipole operator separates into manifestly isoscalar and isovector parts, as
\begin{equation}
\label{eqn-D-alpha-terms}
\Dvec_\alpha=
\tfrac12
\underbrace{\biggl[\sum_i \xvec_i \biggr]}_{\Dvec}
+\tfrac12\alpha
\underbrace{\biggl[\sum_i \xvec_i \tau_{i0}\biggr]}_{\Dveciv}.
\end{equation}
That is, $\Dvec_\alpha=\tfrac12\Dvec+\tfrac12\alpha\Dveciv$, where $\Dvec$ is
simply the \textit{isoscalar} dipole operator from~(\ref{eqn-D-obo}), while
\begin{equation}
\label{eqn-Div-obo}
\Dveciv=U[\xvec\tau_0]=\sum_i \xvec_i \tau_{i0}
\end{equation}
is the \textit{isovector} dipole operator. Since each term is manifestly the
$T_z=0$ spherical tensor component of an isovector operator, this operator is itself, more
precisely, the $T_z=0$ component of an isovector operator.

We focus now on the intrinsic formulation $\Dveciv'$ of the isovector dipole
operator.  We have already found, in~(\ref{eqn-D-intr}), that the intrinsic
isoscalar dipole operator vanishes, so~(\ref{eqn-D-alpha-terms}) implies that
the intrinsic proton or neutron dipole operators $\Dvec_\alpha'$ are simply
proportional to $\Dveciv'$:
\begin{equation}
\label{eqn-Dalpha-intr}
\Dvec_\alpha'=\tfrac12\alpha\Dveciv'.
\end{equation}
Again substituting
$\xvec_i\rightarrow\xvec_i'$~(\ref{eqn-coords-intrinsic}) to obtain the intrinsic operator, we have
\begin{equation}
\label{eqn-Div-intr-subst}
\Dveciv'=\sum_i \bigl(\xvec_i-\frac{1}{A}\sum_j\xvec_j\bigr) \tau_{i0}.
\end{equation}
Observe that this expression for the intrinsic isovector dipole operator
involves double sums over particle indices.  The intrinsic operator is no longer
purely a one-body operator but rather also includes two-body contributions.

Multiplying out the product of sums in~(\ref{eqn-Div-intr-subst}) leaves us with an expression involving a double sum, as
\begin{equation}
\label{eqn-Div-intr-expanded}
\Dveciv'=\sum_i \xvec_i  \tau_{i0} -\frac{1}{A}\sum_{ij} \xvec_j \tau_{i0}.
\end{equation}
The
first term we recognize as simply the original, uncorrected one-body operator
$\Dveciv$ from~(\ref{eqn-Div-obo}),
while the coordinate dependence of the remaining term involves only the center-of-mass degree of
freedom:
\begin{equation}
\label{eqn-Div-intr-sep}
\underbrace{
\Dveciv'
}_{\text{Intrinsic}}
=
\underbrace{
\sum_i \xvec_i  \tau_{i0}
}_{\text{One-body}}
-
\underbrace{
\frac{1}{A}\bigl(\sum_i  \tau_{i0})\bigl(\sum_j\xvec_j)
}_{\text{Center-of-mass}}
.
\end{equation}
Note that the isospin sum in the second term reduces to a simple counting factor
($\sum_i \tau_{i0} = Z-N$, where $Z$ is the proton number, and $N$ is the
neutron number), and we recognize that $\sum_j\xvec_j$ is
simply $\Dvec\propto\xveccm$ from~(\ref{eqn-D-obo}), giving
\begin{equation}
\label{eqn-Div-intr}
\begin{aligned}
\Dveciv'=\Dveciv -\frac{Z-N}{A}\Dvec.
\end{aligned}
\end{equation}

Thus, the intrinsic operator so happens, in this particular case, to decompose
into two separate terms: one representing the original ``naive'' one-body
operator, and the other a c.m.\ (or ``recoil'') contribution.  We highlight this
separation~(\ref{eqn-Div-intr}), since a similar separation occurs below for the
bilinear operators (Sec.~\ref{sec-bilinear}).  While for the dipole operator
considered here in~(\ref{eqn-Div-intr}) both terms are still simply one-body
operators, for the bilinear operators considered below the c.m.\ contribution is
a two-body operator.

As noted in Sec.~\ref{sec-background}, for NCCI calculations in the traditional
$\Nmax$-truncated oscillator basis, the c.m.\ motion in the calculated many-body
wave function is known to exactly factorize from the intrinsic motion, and the
c.m.\ motion of the nuclear many-body state takes on a pure oscillator $0s$ wave
function.  Since the $\Dvec$ operator acts only on the c.m.\ degree of freedom,
its matrix elements are then those of a vector operator between states of zero
angular momentum, which vanish identically [$\trme{0s}{\Dvec}{0s}=0$] by the
angular momentum selection rule.  The matrix elements of the remaining terms
$\Dveciv$ and $\Dveciv'$ in~(\ref{eqn-Div-intr}) must therefore be equal.  The
analogous relation for the bilinear operators considered below plays a
central role in practical calculations of matrix elements of, \textit{e.g.},
the $M1$ and $E2$ operators, between NCCI many-body eigenstates.

However, in more general varieties of many-body calculation, it cannot be
assumed that the c.m.\ wave function separates or has such a simple form.  We
must therefore evaluate transition matrix elements of the full, two-body
intrinsic isovector dipole operator between many-body wave functions.

Before we can evaluate two-body matrix elements, it is first necessary to
represent the operator in the canonical form $V[v]$ for a two-body operator,
defined in~(\ref{eqn-tbo}), so that the operator $v$ on the two-body space can
be identified.  We thus return to the expression for $\Dveciv'$
in~(\ref{eqn-Div-intr-expanded}) and recast it manifestly in the form
of~(\ref{eqn-tbo}).  We must eliminate the one-body term (single sum), but first
eliminate the diagonal ($i=j$) terms from the double sum, and furthermore ensure
that the summand in the double sum is symmetric in the particle
indices, as required of $v_{ij}$.  These tasks are accomplished by applying the
summation identities from Appendix~\ref{sec-app-sum},
successively~(\ref{eqn-diagonal-sum-symm}) and~(\ref{eqn-upgrade-sum}), yielding
\begin{maybemultline}
\label{eqn-Div-intr-sum-foiled}
\Dveciv'=
\frac{1}{2A}\sumprime_{ij} \bigl( \xvec_i \tau_{i0} + \xvec_j \tau_{j0} \bigr)
\ifproofpre{\\}{}
-
\frac{1}{2A}\sumprime_{ij} \bigl( \xvec_i \tau_{j0} + \xvec_j \tau_{i0} \bigr).
\end{maybemultline}
Within each of these two sums, the summand is manifestly symmetric under interchange of
particle indices $i\leftrightarrow j$, so we have indeed obtained a two-body
operator in~(\ref{eqn-Div-intr-sum-foiled}).

Moreover, $\Dveciv'$ is
constructed as an
intrinsic operator, and thus Galilean-invariant, so recall
(Sec.~\ref{sec-nbody}) that we expect it to, more specifically, be a
\textit{relative} two-body operator, and the corresponding
operator $v_{12}$ on the two-particle space should have a coordinate dependence
which involves only the relative coordinate
degrees of freedom.  This is not obvious from~(\ref{eqn-Div-intr-sum-foiled}),
but refactoring yields
\begin{equation}
\Dveciv' =
\frac{1}{2A}\sumprime_{ij} \bigl( \xvec_i -\xvec_j \bigr) \bigl( \tau_{i0} - \tau_{j0} \bigr),
\end{equation}
with the appropriate coordinate dependence of the form $\xvec_i-\xvec_j$.

To explicitly recognize $\Dveciv'$ as a two-body operator,
following~(\ref{eqn-tbo}), we write
\begin{equation}
\label{eqn-Div-intr-formal}
\Dveciv' =
\frac{2}{A}V\biggl[\tfrac12 \bigl( \xvec_1 -\xvec_2 \bigr) \bigl( \tau_{10} - \tau_{20}
  \bigr) \biggr].
\end{equation}
It is natural to pull all $A$ dependence outside of the definition of the
two-body operator $V[\cdots]$ appearing in~(\ref{eqn-Div-intr-formal}), so that
this two-body operator can be defined independent of the number of nucleons $A$
in the many-body problem, and, for practical many-body calculations, its
two-body matrix elements can therefore be evaluated independent of the number of
particles $A$ targeted in the many-body calculation.  Rather, we need only
evaluate the matrix elements of the relative operator
\begin{equation}
\label{eqn-Div-rel}
\Dvecivrel =
\tfrac12 ( \xvec_1 -\xvec_2 ) ( \tau_{10} - \tau_{20} ),
\end{equation}
defined on the two-particle space.  Matrix elements of the corresponding two-body
operator $V[\Dvecivrel]$ on the many-particle space then follow as usual by the
machinery of second quantization.  The two-body matrix elements of the relative
operator $\Dvecivrel$ must then simply be scaled by the ``counting factor''
$2/A$ to give two-body matrix elements for the intrinsic operator $\Dveciv'$.

Note that we can explicitly represent this relative operator $\Dvecivrel$ in terms of the
relative coordinate $\xvecrel$, but we must take care that the expression
depends on the choice of convention in the definition~(\ref{eqn-coords-rcm}) of
$\xvecrel$:
\begin{equation}
\label{eqn-Div-rel-rcm}
\Dvecivrel =
\tBstack{       1}{\sqrt2} \tfrac12 \xvecrel ( \tau_{10} - \tau_{20}).
\end{equation}
Also, note that the isospin dependence $ \tau_{10} - \tau_{20}$ enters as the
zero component of the isovector operator $\tauvec_{1} - \tauvec_{2}$, justifying
the identification of $\Dvecivrel$, and hence $\Dveciv$, as indeed an isovector
operator.  The relevant isospin selection rules and relations needed to evaluate
isospin-reduced two-body matrix elements of this operator are provided for
reference in Appendix~\ref{sec-app-isospin}.


\section{Relations for bilinear operators}
\label{sec-bilinear}

The essential input for working with a two-body operator in a
many-body calculation is its two-body matrix elements.  As illustrated for the familiar case of the intrinsic kinetic energy in Sec.~\ref{sec-background}, two main approaches
exist for evaluating these matrix elements.  These require reexpressing the operator, relative to its defining
form in terms of $\xvec'_i$ and $\pvec'_i$, either: (1)~as an explicit two-body
operator, that is, expressed in terms of the two-body system's relative
coordinate and momentum, or (2)~decomposed into one-body and separable two-body
terms.  We estabish the necessary relations for the two-body
operator obtained by transforming a bilinear operator to the intrinsic frame, first for an isoscalar bilinear operator
(Sec.~\ref{sec-bilinear-isoscalar}), then for an isovector bilinear operator (Sec.~\ref{sec-bilinear-isovector}).


\subsection{Isoscalar bilinear operators}
\label{sec-bilinear-isoscalar}

We now consider the problem of transforming a ``bilinear'' spherical tensor
operator to the intrinsic frame.  That is, we start with the one-body operator
obtained from a spherical tensor product of the form
$(\etavec\times\xivec)_{L_0}$, where $\etavec$ and $\xivec$ represent either
coordinate and/or momentum vectors or oscillator creation and/or annihilation
operators.  These latter (ladder) operators are defined and their properties
reviewed in Appendix~\ref{sec-app-ladder}.
We are thus considering a one-body operator of
the form
\begin{equation}
\label{eqn-T-defn}
\Top= U[(\etavec\times\xivec)_{L_0}]=\sum_i(\etavec_i\times\xivec_i)_{L_0}.
\end{equation}
As the operators $\etavec$ and $\xivec$ are spherical tensor operators of rank
$1$, \textit{i.e.}, vector operators, their angular momenta can couple to give a
total angular momentum $L_0=0$, $1$, or $2$.

All such bilinear operators have the same structure, as far as their
transformation to the intrinsic frame is concerned.  We can therefore treat all
such operators generically in a single, generic derivation.  This affords not
only a certain efficiency of effort but also highlights the otherwise nonobvious
parallels (and distinctions) among a variety of structurally similar operators.
As already motivated in the introduction (Sec.~\ref{sec-intro}), depending on
the nature of the basis for the problem, we may find it necessary to recast the
bilinear operator into one of two forms: (1)~a one body contribution plus a
\textit{separable} two-body operator or (2)~a \textit{relative} two-body
operator.

Regardless of whether $\etavec$ and $\xivec$ represent $\xvec$, $\kvec$,
$\cvec^\dagger$, or $\cvec$, the transformation to the intrinsic frame, in
either~(\ref{eqn-coords-intrinsic}) for the coordinates and momenta
or~(\ref{eqn-ladder-intrinsic}) for the ladder operators, is of the form:
\begin{equation}
\label{eqn-etap-xip-defn}
\etavec'_i = \etavec_i - \frac1A\sum_j\etavec_j
\qquad
\xivec'_i = \xivec_i - \frac1A\sum_j\xivec_j.
\end{equation}
The intrinsic operator $\Top'$ obtained from $\Top$,
is then defined by
\begin{equation}
\label{eqn-Tp-implicit}
\Top'=\sum_i(\etavec'_i\times\xivec'_i)_{L_0},
\end{equation}
and thus has the form
\begin{equation}
\label{eqn-Tp-sum}
\Top'=\sum_i\Bigl[
\bigl(\etavec_i- \frac1A\sum_j\etavec_j\bigr)
 \times
\bigl(\xivec_i- \frac1A\sum_k\xivec_k\bigr)
\Bigr]_{L_0}.
\end{equation}
When multiplied out, this expression for the intrinsic operator involves double
and even triple sums over particle indices:
\begin{multline}
\label{eqn-Tp-foiled}
\Top'=
\sum_i \bigl(\etavec_i\times\xivec_i\bigr)_{L_0}
-\frac1A \sum_{ik} \bigl(\etavec_i \times \xivec_k\bigr)_{L_0}
\ifproofpre{\\}{}
-\frac1A \sum_{ij} \bigl(\etavec_j \times \xivec_i\bigr)_{L_0}
+\frac1{A^2} \sum_{ijk} \bigl(\etavec_j \times \xivec_k\bigr)_{L_0}.
\end{multline}
The intrinsic operator is no longer purely a one-body operator.  However, this
expression involves at most two-body, not three-body, contributions: the triple
sum $\propto\sum_{ijk}(\etavec_j\times\xivec_k)_{L_0}$ in the final term
involves summation over an index $i$ not appearing in the summand, which simply
yields a counting factor (recall $\sum_i1=A$).

First, we examine the structure of the expression in~(\ref{eqn-Tp-foiled}).
Renaming summation indices (and recognizing $\sum_i1=A$) allows us to collect like terms, leaving only two
distinct terms:
\begin{equation}
\label{eqn-Tp-foiled-again}
\Top'=
\sum_i \bigl(\etavec_i\times\xivec_i\bigr)_{L_0}
-\frac1A \sum_{ij} \bigl(\etavec_i \times \xivec_j\bigr)_{L_0}.
\end{equation}
The first term we recognize as simply the original, uncorrected one-body
operator $\Top$, while the remaining term can be factored into parts which act
only on the center-of-mass degree of freedom:
\begin{equation}
\label{eqn-Tp-separated}
\underbrace{\Top'}_{\text{Intrinsic}}=
\underbrace{\sum_i \bigl(\etavec_i\times\xivec_i\bigr)_{L_0}}_{\text{One-body}}
-\underbrace{
\frac1A  \Bigl[\bigl(\sum_{i}\etavec_i\bigr) \times \bigl(
  \sum_{j}\xivec_j\bigr)\Bigr]_{L_0}
}_{\text{Center-of-mass}}.
\end{equation}
Thus, as with the dipole operator in~(\ref{eqn-Div-intr-sep}), the intrinsic
operator conveniently separates into the ``naive'' one-body operator and a
c.m.\ recoil contribution.

The practical significance of this separation~(\ref{eqn-Tp-separated}) similarly
arises when we set out to evaluate matrix elements of the intrinsic operator
between wave functions obtained from solving the many-body problem,
\textit{e.g.}, for transition matrix elements, moments, or expectation values.
Again, for NCCI calculations in the traditional $\Nmax$-truncated oscillator
basis, the c.m.\ motion factorizes and is described by an oscillator $0s$ wave
function.

For the case where the bilinear operator is nonscalar ($L_0>0$), the
c.m.\ contribution $\propto\trme{0s}{(\etaveccm\times\xiveccm)_{L_0}}{0s}$
vanishes identically, by the angular momentum triangle selection rule.  Then
matrix elements of the naive one-body operator may again be evaluated in place
of matrix elements of the more computationally demanding two-body intrinsic
operator.  Thus, \textit{e.g.}, the naive one-body quadrupole operator is
commonly used in place of the intrinsic quadrupole operator in $\Nmax$-truncated
oscillator-basis NCCI calculations. (So far we have strictly only justified this
for the isoscalar, or mass, quadrupole operator, while the less obvious case of
the electric quadrupole operator is addressed below in
Sec.~\ref{sec-bilinear-isovector}.)  However, this substitution is no longer
justified when other bases are used and the factorized $0s$ c.m.\ wave function
is sacrificed.

For a scalar bilinear operator ($L_0=0$), no such angular momentum selection rule
applies, and the c.m.\ correction to a matrix element is in general nonzero.
Thus, \textit{e.g.}, in evaluating the r.m.s.\ radius, it is necessary to
calculate the expectation value of the intrinsic squared radius operator, which
is a rotational scalar. The expectation value of the naive one-body operator
does not simply equal that of the intrinsic operator.  Nonetheless, the
c.m.\ correction reduces to the expectation value of
$\rcm^2=\xveccm\cdot\xveccm$ in a $0s$ oscillator state, which is analytically
known from simple application of harmonic oscillator laddering relations
(\textit{e.g.}, Ref.~\cite{cockrell2012:li-ncfc}).  Thus, when working in an
$\Nmax$-truncated oscillator basis, the one-body operator can be used in place
of the intrinsic operator by application of a simple correction.  For other bases,
the full two-body intrinsic operator must, in general, be used.

Next, we pursue the representation of $\Top'$ as a one-body contribution plus a
separable two-body operator.  Although the expression~(\ref{eqn-Tp-foiled-again}) involves both
single and double sums over particle indices, these terms do not directly
represent the one-body and two-body parts.  To bring the double sums into the
form~(\ref{eqn-tbo}) required for a two-body operator, recall that we must
extract the diagonal terms of the sum [via~(\ref{eqn-diagonal-sum})], leaving behind
a restricted sum ($\tsumprime_{ij}$), and we must also ensure that the summand
is symmetrized with respect to interchange of particle indices ($i\leftrightarrow j$):
\begin{maybemultline}
\label{eqn-Tp-obo-separable-tbo-sum}
\Top'=
\Bigl(1-\frac{1}{A}\Bigr)
\sum_i \bigl(\etavec_i\times\xivec_i\bigr)_{L_0}
\ifproofpre{\\}{}
-\frac1{2A} \sumprime_{ij} \Bigl[
 \bigl(\etavec_i \times \xivec_j\bigr)_{L_0}
+
 \bigl(\etavec_j \times \xivec_i\bigr)_{L_0}
\Bigr].
\end{maybemultline}
We recognize the first and second terms in this
expression~(\ref{eqn-Tp-obo-separable-tbo-sum}) as representing one-body and
two-body operators, respectively.  Explicitly, in the notation of Sec.~\ref{sec-nbody}:
\begin{maybemultline}
\label{eqn-Tp-obo-separable-tbo-formal}
\Top'=
\Bigl(1-\frac{1}{A}\Bigr)
U\Bigl[
 \bigl(\etavec\times\xivec\bigr)_{L_0}
 \Bigr]
\ifproofpre{\\}{}
-\frac1{A}
V\Bigl[
 \bigl(\etavec_1 \times \xivec_2\bigr)_{L_0}
+
 \bigl(\etavec_2 \times \xivec_1\bigr)_{L_0}
\Bigr].
\end{maybemultline}
Note that each term $\bigl(\etavec_1 \times \xivec_2\bigr)_{L_0}$ or
$\bigl(\etavec_2 \times \xivec_1\bigr)_{L_0}$ in the two-body operator is the
product of a factor acting only on the first particle and a factor acting only
on the second particle.  This separable form permits the use of Racah's
reduction formula to evaluate two-body reduced matrix elements, in terms of the
much more easily computed reduced matrix elements of $\etavec$ and $\xivec$ in
the single-particle basis.

Alternatively, we recast $\Top'$ as a relative two-body operator.  To do so,
recall from our treatment of the dipole operator (Sec.~\ref{sec-dipole}) we must
now upgrade the one-body contribution so that it is represented as a two-body
operator [via~(\ref{eqn-upgrade-sum})]:
\begin{multline}
\label{eqn-Tp-tbo-sum-uncombined}
\Top'=
\frac{1}{2A} \sumprime_{ij} \Bigl[
 \bigl(\etavec_i \times \xivec_i\bigr)_{L_0}
+
 \bigl(\etavec_j \times \xivec_j\bigr)_{L_0}
 \Bigr]
\ifproofpre{\\}{}
-\frac1{2A} \sumprime_{ij} \Bigl[
 \bigl(\etavec_i \times \xivec_j\bigr)_{L_0}
+
 \bigl(\etavec_j \times \xivec_i\bigr)_{L_0}
\Bigr].
\end{multline}
While each term here is symmetric under interchange of particle indices, it is
still not immediately apparent that the two-body operator involves only the
relative coordinate degree of freedom.  We must
refactor~(\ref{eqn-Tp-tbo-sum-uncombined}) to obtain
\begin{equation}
\label{eqn-Tp-tbo-sum-combined}
\Top'=
\frac{1}{2A} \sumprime_{ij}
\bigl[
 \bigl(\etavec_i-\etavec_j\bigr)
\times
 \bigl(\xivec_i-\xivec_j\bigr)
\bigr]_{L_0}.
\end{equation}
Explicitly recognizing this expression as representing a relative two-body operator, in
the notation of~(\ref{eqn-tbo}), we have
\begin{equation}
\label{eqn-Tp-tbo-formal}
\Top'=
\frac{2}{A}
\,
V \Bigl[
\underbrace{
\tfrac12
\bigl[
 \bigl(\etavec_1-\etavec_2\bigr)
\times
 \bigl(\xivec_1-\xivec_2\bigr)
\bigr]_{L_0}
}_{\Toprel}
\Bigr].
\end{equation}

As with the dipole operator in~(\ref{eqn-Div-intr-formal}), we have extracted the
$A$ dependence from the definition of the two-body operator $V[\cdots]$
appearing in~(\ref{eqn-Tp-tbo-formal}), and have extracted an additional factor
of $2$ to ensure that the intrinsic operator $\Top'$ simply reduces to the relative operator
\begin{equation}
\label{eqn-Trel}
\Toprel=
\tfrac12
\bigl[
 \bigl(\etavec_1-\etavec_2\bigr)
\times
 \bigl(\xivec_1-\xivec_2\bigr)
\bigr]_{L_0}
\end{equation}
appearing inside the $V[\cdots]$ when evaluated on the two-particle space.  We need
only evaluate the $A$-independent matrix elements of this relative operator on
the two-particle space.  Then, as usual, matrix elements of $V[\Toprel]$ on the
many-particle space follow by second quantization.  These must then be scaled by the
counting factor $2/A$ to yield matrix elements for the full intrinsic
operator $\Top'$.


\subsection{Isovector bilinear operators}
\label{sec-bilinear-isovector}

Electromagnetic operators of bilinear type are not of the simple isoscalar
form~(\ref{eqn-T-defn}) but rather distinguish protons and neutrons.  As with the
dipole operator in Sec.~\ref{sec-dipole}, rather than restricting the summation
to just protons or just neutrons, which would sacrifice the
simplicity of working with a true one-body operator, we instead modify the
operator on the single-particle space so that it selects either for protons or for
neutrons.  Thus, we have proton and neutron bilinear operators
\begin{equation}
\label{eqn-T-proton-neutron-sum}
\begin{aligned}
T_{p,L_0}&=\sum_i(\etavec_i\times\xivec_i)_{L_0}\delta_{i,p}\\
T_{n,L_0}&=\sum_i(\etavec_i\times\xivec_i)_{L_0}\delta_{i,n},
\end{aligned}
\end{equation}
where, again, $\delta_{i,p}=1$ if the $i$th particle is a proton and
$\delta_{i,p}=0$ if the $i$th particle is a neutron, and \textit{vice versa} for
$\delta_{i,n}$.

Although the one-body operator $T_{p,L_0}$ acts only upon the protons (and
similarly $T_{n,L_0}$ acts only upon the neutrons), note that this is no longer
true for the corresponding intrinsic operator.  Taking, for example, the
intrinsic proton operator
\begin{equation}
\label{eqn-Tp-proton-sum}
T'_{p,L_0}=\sum_i\Bigl[
\bigl(\etavec_i- \frac1A\sum_j\etavec_j\bigr)
 \times
\bigl(\xivec_i- \frac1A\sum_k\xivec_k\bigr)
\Bigr]_{L_0}\delta_{i,p},
\end{equation}
neutrons contribute through the c.m.\ recoil corrections.  That is, while the
factor $\delta_{i,p}$ ensures that only values of the particle index $i$ which
correspond to protons contribute, there is no such restriction on the values of
the particle indices $j$ and $k$ which contribute.\footnote{The proton and
  neutron intrinsic bilinear operators are thus substantially different in
  nature from proton-only or neutron-only interaction two-body operators, most
  notably the Coulomb interaction, which genuinely only acts on protons.  These
  operators have a different isospin structure, involving not only isoscalar and
  isovector but also isotensor contributions.}

As with the dipole operator in Sec.~\ref{sec-dipole}, it is convenient to
consolidate the treatment of the proton and neutron bilinear operators into an
isovector bilinear operator.  We first use
the selection operator $\delta_\alpha$, defined in (\ref{eqn-delta-alpha}), to define
\begin{equation}
\label{eqn-T-alpha-sum}
\Topalpha=\sum_i(\etavec_i\times\xivec_i)_{L_0}\delta_{i,\alpha},
\end{equation}
where the proton operator is obtained for $\alpha=+1$, and the neutron
operator for $\alpha=-1$.
The proton or neutron operator then separates into isoscalar and isovector parts as
\begin{equation}
\label{eqn-T-alpha-terms}
\Topalpha=
\tfrac12
\underbrace{\biggl[\sum_i(\etavec_i\times\xivec_i)_{L_0}\biggr]}_{\Top}
+\alpha\tfrac12
\underbrace{\biggl[\sum_i(\etavec_i\times\xivec_i)_{L_0}\tau_{i0}\biggr]}_{\Topiv}.
\end{equation}
The first term simply involves the isoscalar operator $\Top$
of~(\ref{eqn-T-defn}), while the second term defines the isovector operator
\begin{equation}
\label{eqn-Tiv-defn}
\Topiv= \sum_i(\etavec_i\times\xivec_i)_{L_0}\tau_{i0}.
\end{equation}

The intrinsic form of this isovector bilinear operator, obtained by
the substitution~(\ref{eqn-etap-xip-defn}), is
\begin{equation}
\label{eqn-Tivp-sum}
\Topiv'=\sum_i\Bigl[
\bigl(\etavec_i- \frac1A\sum_j\etavec_j\bigr)
 \times
\bigl(\xivec_i- \frac1A\sum_k\xivec_k\bigr)
\Bigr]_{L_0}\tau_{i0}.
\end{equation}
Multiplying out the product again yields terms involving summations over one or
more particle indices, as in~(\ref{eqn-Tp-foiled}), but now also involving the isospin
factor of $\tau_{i0}$:
\begin{multline}
\label{eqn-Tivp-foiled}
\Topiv'=
\sum_i \bigl(\etavec_i\times\xivec_i\bigr)_{L_0} \tau_{i0}
\\
-\frac1A \sum_{ik} \bigl((\etavec_i \tau_{i0}) \times \xivec_k\bigr)_{L_0}
-\frac1A \sum_{ij} \bigl(\etavec_j \times (\xivec_i\tau_{i0})\bigr)_{L_0}
\\
+\frac1{A^2} \sum_{ijk} \tau_{i0} \bigl(\etavec_j \times \xivec_k\bigr)_{L_0}.
\end{multline}
The operator $\tau_{i0}$, although isovector, is a rotational scalar, and may
thus be moved around within the products without regard for the angular momentum
coulpling: here, in~(\ref{eqn-Tivp-foiled}), we group it with the other factors
involving the particle index~$i$, to emphasize this connection.  In the triple
sum appearing in the final term, $\propto\sum_{ijk} \tau_{i0} \bigl(\etavec_j
\times \xivec_k\bigr)_{L_0}$, the summation over $i$ simply yields the counting
factor $\sum_i \tau_{i0} = Z-N$, as in~(\ref{eqn-Div-intr}).

However, the factors of $\tau_{i0}$ mean that, even after renaming particle
indices, there are no longer any strictly identical terms to combine as
in~(\ref{eqn-Tp-foiled-again}).  Rather, we can arrange the expression into
three sums involving different combinations of summation indices and
$\tau_{i0}$ dependences:
\begin{multline}
\label{eqn-Tivp-foiled-again}
\Topiv'=
\sum_i \bigl(\etavec_i\times\xivec_i\bigr)_{L_0} \tau_{i0}
\\
-\frac1A \sum_{ij}
  \Bigl[
    \bigl((\etavec_i \tau_{i0}) \times \xivec_j\bigr)_{L_0}
   +\bigl(\etavec_j \times (\xivec_i\tau_{i0})\bigr)_{L_0}
  \Bigr]
\\
+\frac{1}{A}\Bigl(\frac{Z-N}{A}\Bigr)
\sum_{ij} \bigl(\etavec_i \times \xivec_j\bigr)_{L_0}.
\end{multline}
The first term of~(\ref{eqn-Tivp-foiled-again}) is recognized as the naive
one-body operator $\Topiv$ of~(\ref{eqn-Tiv-defn}).  The final term is again
purely a c.m.\ term $\propto (\etaveccm \times \xiveccm)_{L_0}$.  However, the
isospin factors in the middle terms of~(\ref{eqn-Tivp-foiled-again}) preclude
the separation of $\Topiv'$ into one-body and c.m.\ terms, in the simple way
obtained above for the isoscalar bilinear operator in the corresponding
expression~(\ref{eqn-Tp-separated}).
The question thus arises as to whether or not, when the c.m.\ wave function has
known $0s$ form, the naive one-body operator can still be used in place of the
full intrinsic operator~--- and, if not, what is the expression for the
c.m.\ correction?

The middle terms of~(\ref{eqn-Tivp-foiled-again}), involving $\sum_{ij} \bigl((\etavec_i
\tau_{i0}) \times \xivec_j\bigr)_{L_0}$ and $\sum_{ij}\bigl(\etavec_j \times
(\xivec_i\tau_{i0})\bigr)_{L_0}$ can most simply be addressed by leveraging the
results for the dipole operator obtained in Sec.~\ref{sec-dipole} to represent
them in terms of intrinsic and c.m.\ operators to which selection rules can be
applied.  While the dipole operator was defined in Sec.~\ref{sec-dipole} in terms of the spatial
coordinate $\xvec$, we may generalize this definition to a general coordinate, momentum, or ladder operator
$\xivec$, letting
\begin{math}
\Dvec^{\xivec}=\sum_i\xivec_i,
\end{math}
analogous to~(\ref{eqn-D-obo}), and
\begin{math}
\Dveciv^{\xivec}=\sum_i \xivec_i \tau_{i0},
\end{math}
analogous to~(\ref{eqn-Div-obo}).  Then, $\Dvec^{\xivec}$ is again a pure
c.m.\ operator, and we have the relation
\begin{math}
\Dveciv^{\xivec\,\prime}=\Dveciv^{\xivec} -[(Z-N)/A]\Dvec^{\xivec},
\end{math}
analogous to~(\ref{eqn-Div-intr}).
Thus, taking the first of the middle terms in~(\ref{eqn-Tivp-foiled-again}), we recognize
\begin{multline}
\sum_{ij} \bigl((\etavec_i \tau_{i0}) \times \xivec_j\bigr)_{L_0}
=(\Dveciv^{\etavec}\times\Dvec^{\xivec})_{L_0}
\ifproofpre{\\}{}
=(\Dveciv^{\etavec\,\prime}\times\Dvec^{\xivec})_{L_0}+\Bigl(\frac{Z-N}{A}\Bigr)(\Dvec^{\etavec}\times\Dvec^{\xivec})_{L_0}.
\end{multline}
The matrix element of the first term in this expression vanishes by the angular momentum triangle
selection rule, which enforces vanishing $\trme{0s}{\Dvec^{\xivec}}{0s}$, while the second
term of this expression is a pure c.m.\ operator of the same form as the last term
of~(\ref{eqn-Tivp-foiled-again}).  The second of the middle terms
in~(\ref{eqn-Tivp-foiled-again}) may be handled similarly.  Thus, the matrix
element of $\Topiv'$ between states with $0s$ center-of-mass motion again
separates into the naive one-body contribution plus a pure center-of-mass
contribution, which again vanishes by angular momentum selection if $L_0\neq0$.

Keeping in mind our first aim, to separate $\Topiv'$ into one-body and separable
two-body parts, we extract the diagonal terms from the double sums
in~(\ref{eqn-Tivp-foiled-again}) [via~(\ref{eqn-diagonal-sum})], to
obtain the desired restricted sum ($\tsumprime_{ij}$), and
explicitly symmetrize the expressions under interchange of particle indices
($i\leftrightarrow j$), obtaining
\begin{multline}
\label{eqn-Tivp-obo-separable-tbo-sum}
\Topiv'
=
\ifproofpre{\\}{}
\Bigl(1-\frac{2}{A}\Bigr)
\sum_i \bigl(\etavec_i\times\xivec_i\bigr)_{L_0} \tau_{i0}
+\frac{1}{A}\Bigl(\frac{Z-N}{A}\Bigr)
\sum_i \bigl(\etavec_i\times\xivec_i\bigr)_{L_0}
\\
-\frac1{2A} \sumprime_{ij} \Bigl[
 \bigl(\etavec_i \times \xivec_j\bigr)_{L_0}
+
 \bigl(\etavec_j \times \xivec_i\bigr)_{L_0}
\Bigr]
(\tau_{i0}+\tau_{j0})
\\
+\frac{1}{2A}\Bigl(\frac{Z-N}{A}\Bigr)
\sumprime_{ij} \Bigl[
 \bigl(\etavec_i \times \xivec_j\bigr)_{L_0}
+
 \bigl(\etavec_j \times \xivec_i\bigr)_{L_0}
\Bigr]
.
\end{multline}
Explicitly recognizing the first two terms as one-body operators and the latter
two terms as two-body operators, we have
\begin{multline}
\label{eqn-Tivp-obo-separable-tbo-formal}
\Topiv'=
\ifproofpre{\\}{}
\Bigl(1-\frac{2}{A}\Bigr)
U\Bigl[
 \bigl(\etavec\times\xivec\bigr)_{L_0} \tau_0
\Bigr]
+\frac{1}{A}\Bigl(\frac{Z-N}{A}\Bigr)
U\Bigl[
 \bigl(\etavec\times\xivec\bigr)_{L_0}
\Bigr]
\\
-\frac1{A}
V\Bigl[
\bigl[
 \bigl(\etavec_1 \times \xivec_2\bigr)_{L_0}
+
 \bigl(\etavec_2 \times \xivec_1\bigr)_{L_0}
\bigr] (\tau_{10}+\tau_{20})
\Bigr]
\\
-\frac{1}{A}\Bigl(\frac{Z-N}{A}\Bigr)
V\Bigl[
 \bigl(\etavec_1 \times \xivec_2\bigr)_{L_0}
+
 \bigl(\etavec_2 \times \xivec_1\bigr)_{L_0}
\Bigr].
\end{multline}
The second and fourth terms in~(\ref{eqn-Tivp-obo-separable-tbo-formal}) involve
the one-body and separable two-body operators which already appear in the
analogous decomposition~(\ref{eqn-Tp-obo-separable-tbo-formal}) for the
isoscalar operator $\Top'$, although now with different, $(Z-N)$-dependent
coefficients.  The first and third terms involve new, isovector one-body and
two-body operators. Note that $\tauvec_{1}+\tauvec_{2}=2\vec{T}$ is simply
proportional to the total isospin operator $\vec{T}$ for the two-body system, so
the combination $\tau_{10}+\tau_{20}$ appearing in the third term
of~(\ref{eqn-Tivp-obo-separable-tbo-formal}) is simply $2T_0$.

Alternatively, to recast $\Topiv'$ as a relative two-body operator, we must upgrade
the one-body contribution in~(\ref{eqn-Tivp-obo-separable-tbo-sum}) so that it
is represented as a two-body operator [via~(\ref{eqn-upgrade-sum})].  However, the
algebra is considerably more involved than for the isoscalar operator $\Top'$
in~(\ref{eqn-Tp-tbo-sum-uncombined}) above, and we omit the
details.  [We only note that, in order to combine like terms, it is helpful to be
able to introduce or remove an explicit dependence on the
isospin operators $\tau_{i0}$ and $\tau_{j0}$ within the sums, via the identity
\begin{math}
(Z-N)\tsumprime_{ij}(u_i+u_j)
=
\tsumprime_{ij}(u_i\tau_{i0}+u_j\tau_{j0})
+(A-1)
\tsumprime_{ij}(u_i\tau_{j0}+u_j\tau_{i0}),
\end{math}
which may be obtained by combining~(\ref{eqn-upgrade-sum}) with the identity $\sum_i \tau_{i0}=Z-N$.]
In the end, we obtain an expression for $\Topiv'$ involving sums of the desired relative
two-body form:
\begin{multline}
\label{eqn-Tivp-tbo-sum-combined}
\Topiv'=
\frac{1}{2A} \sumprime_{ij} \Bigl[
\bigl[
 \bigl(\etavec_i-\etavec_j\bigr)
\times
 \bigl(\xivec_i-\xivec_j\bigr)
\bigr]_{L_0}(\tau_{i0}+\tau_{j0})
\Bigr]
\\
-\frac{1}{2A}\Bigl(\frac{Z-N}{A}\Bigr)
\sumprime_{ij}
\bigl[
 \bigl(\etavec_i-\etavec_j\bigr)
\times
 \bigl(\xivec_i-\xivec_j\bigr)
\bigr]_{L_0}
.
\end{multline}
Explicitly recognizing both sums in~(\ref{eqn-Tivp-tbo-sum-combined}) as
two-body operators, we have
\begin{multline}
  \label{eqn-Tivp-tbo-sum-formal}
  \Topiv'=
  \frac{2}{A}
  \,
  V \Bigl[
    \underbrace{
    \overbrace{
      \tfrac12
      \bigl[
        \bigl(\etavec_1-\etavec_2\bigr)
        \times
        \bigl(\xivec_1-\xivec_2\bigr)
        \bigr]_{L_0}
    }^{\Toprel}
    \,(2T_0)
    }_{\Topivrel}
    \Bigr]
  \\
  -
  \frac{2}{A}\Bigl(\frac{Z-N}{A}\Bigr)
  \,
  V \Bigl[
    \underbrace{
      \tfrac12
      \bigl[
        \bigl(\etavec_1-\etavec_2\bigr)
        \times
        \bigl(\xivec_1-\xivec_2\bigr)
        \bigr]_{L_0}
    }_{\Toprel}
    \Bigr]
  .
\end{multline}
Here we recognize that, as noted above,
$\tauvec_{1}+\tauvec_{2}=\tauvec\equiv2\vec{T}$ is simply proportional to the total isospin
operator on the two-particle space, and so $\tau_{10}+\tau_{20}=2T_0$.

Both two-body operators appearing in~(\ref{eqn-Tivp-tbo-sum-formal}) are
manifestly Galilean-invariant (and thus relative) two-body operators.  They are
expressed in terms of the relative operators $\Topivrel=2\Toprel T_0$ and
$\Toprel$ on the two-particle space, where $\Toprel$ was defined previously
in~(\ref{eqn-Trel}).  Note that, in both terms
of~(\ref{eqn-Tivp-tbo-sum-formal}), as previously in~(\ref{eqn-Tp-tbo-formal}),
we extract a factor of $2$ in the overall coefficient, so that the intrinsic
operator, when evaluated on the two-body system ($A=2$), is more simply related
to the relative operators defined on the two-particle space.

To set up a many-body calculation, we need only evaluate the $A$-independent and
$(Z-N)$-independent matrix elements of these two relative operators on the
two-particle space.  These two-body matrix element may thus be evaluated once
and for all, independent of the number of nucleons in the system.  Then the
two-body matrix elements for the full intrinsic operator $\Topiv'$ on a system
with a given number of nucleons may then be obtained as the linear combination
with the appropriate scale factors, involving $A$ and $Z-N$,
from~(\ref{eqn-Tivp-tbo-sum-formal}).  The two-body matrix elements of $\Toprel$
have already been noted as being required for the isoscalar operator $\Top'$,
via~(\ref{eqn-Tp-tbo-formal}).  Furthermore, the two-body matrix elements of
$\Topivrel$ may be simply obtained from those of $\Toprel$, as given in isospin
scheme in~(\ref{eqn-rme-Tivrel}), due to the clean factorization of $\Topivrel$
into spatial and isospin factors.

With the isoscalar and isovector bilinear operators now addressed, we can, of
course, recover the intrinsic form of the proton or neutron bilinear operators.
We recall the expression~(\ref{eqn-T-alpha-terms}) for $\Topalpha$ in terms of
$\Top$ and $\Topiv$ and combine the intrinsic forms $\Top'$
from~(\ref{eqn-Tp-tbo-formal}) and $\Topiv'$
from~(\ref{eqn-Tivp-tbo-sum-formal}) of these operators, to obtain
\begin{multline}
\label{eqn-Talpha-tbo-sum-formal}
\Topalpha'=
\frac{1}{A}\Bigl(1 - \alpha \frac{Z-N}{A}\Bigr)
\,
V \Bigl[
\tfrac12
\bigl[
 \bigl(\etavec_1-\etavec_2\bigr)
\times
 \bigl(\xivec_1-\xivec_2\bigr)
\bigr]_{L_0}
\Bigr]
\\
+
\alpha\frac{1}{A}
\,
V \Bigl[
\tfrac12
\bigl[
 \bigl(\etavec_1-\etavec_2\bigr)
\times
 \bigl(\xivec_1-\xivec_2\bigr)
\bigr]_{L_0}
\,(2T_0)
\Bigr]
.
\end{multline}


\section{Conclusion}
\label{sec-conclusion}

When solving the nuclear many-body problem, there is a tension between the
simultaneous needs to respect antisymmetry under particle exchange and to
preserve the underlying Galilean invariance of the problem. Typically, for
computational methods which can be applied beyond the very lightest few-nucleon
systems, manifest Galilean invariance is sacrificed in favor of manifest
antisymmetry. However, we still require the ability to compute observables which
are free of center-of-mass contamination.  The consequence is that we must work
with intrinsic operators for observables.

In this work, we have derived generic expressions relating one-body operators
which are bilinear in coordinates and/or momenta to their intrinsic
counterparts. We give expressions directly usable for calculating two-body
matrix elements in the laboratory frame, as well as matrix elements in the
relative coordinate degree of freedom (which yield two-body matrix elements via
the Moshinsky transformation). These formulae are applicable to both scalar and
non-scalar bilinear operators, as well as to both the isoscalar and isovector
forms of these operators. In appendices, we provide further results for
reference in working with and evaluating two-body matrix elements of intrinsic
operators, including the coefficients necessary to realize physical operators of
interest in terms of the generic expressions for bilinear operators.

The derivations of the present expressions, while elementary in their methods,
require care in their execution, in order to be of practical use to nuclear
many-body practitioners.  Our effort to synthesize and explicitly establish
these results is motivated by their expected usefulness in implementing
\textit{ab initio} nuclear many-body calculations, using both traditional and
$\grpsptr\supset\grpu{3}$ symmetry-adapted techniques.




\begin{acknowledgments}
We thank Pieter Maris and James P.~Vary for valuable discussions and Jakub Herko
for a careful reading and detailed comments on the manuscript.  This material is
based upon work supported by the U.S.~Department of Energy, Office of Science,
Office of Nuclear Physics, under Award Number DE-FG02-95ER-40934.  TRIUMF
receives federal funding via a contribution agreement with the National Research
Council of Canada.
\end{acknowledgments}


\appendix


\section{Physical operators realized as bilinear operators}
\label{sec-app-physical}

We have thus far kept our expression for the bilinear operators in the generic
form $(\etavec\times\xivec)_{L_0}$, in both its isoscalar~(\ref{eqn-T-defn}) and
isovector~(\ref{eqn-Tiv-defn}) variants.  This generic form must then, of
course, be specialized to the particular operators, representing either physical
observables or group generators, of interest in nuclear physics applications.
Once one of these operators $O$ is related, as
\begin{equation}
  \label{eqn-O}
  O=\kappa
  U[(\etavec\times\xivec)_{L_0}],
\end{equation}
to the generic bilinear form, then the translation to an explicitly intrinsic
operator (by replacement $\etavec\rightarrow\etavec'$ and
$\xivec\rightarrow\xivec'$) and subsequent decompositions into either separable
or relative two-body forms follow immediately from the results of
Sec.~\ref{sec-bilinear}.

In this appendix, we set out specializations of~(\ref{eqn-O}) to specific
operators of physical interest and highlight some of the practical consequences
of the relations from Sec.~\ref{sec-bilinear} for these operators.  We consider
bilinear operators first involving coordinates and momenta~---
$(\xvec\times\xvec)_{L_0}$, $(\kvec\times\kvec)_{L_0}$, and
$(\xvec\times\kvec)_{L_0}$~--- and then involving oscillator ladder
operators~--- $(\cvec^\dagger\times\cvec)_{L_0}$,
$(\cvec^\dagger\times\cvec^\dagger)_{L_0}$, and $(\cvec\times\cvec)_{L_0}$~---
with $L_0=0,1,2$.  The results are summarized in
Table~\ref{tab-physical-bilinear}.  The isoscalar forms are provided in the
table, while the corresponding proton or neutron forms may be obtained through
the corresponding substitutions in~(\ref{eqn-T-proton-neutron-sum}).

Although the relative operator $\Toprel$ on the two-body space, as defined
in~(\ref{eqn-Trel}), may clearly be expressed in terms of relative quantities
$\etavecrel\propto \etavec_1-\etavec_2$ and $\xivecrel\propto
\xivec_1-\xivec_2$, the proportionality factor depends upon the particular
operators $\etavec$ and $\xivec$ involved.  We must therefore allow for a scale
factor $s$ in
\begin{equation}
\label{eqn-Trel-etarel-xirel}
\Toprel=
s \bigl(\etavecrel\times \xivecrel\bigr)_{L_0}
,
\end{equation}
where $s$ is specified explicitly for the various physical operators in
Table~\ref{tab-physical-bilinear}.
This scale factor is practically
important in the evaluation of two-body matrix elements for $\Top'$, when these
are obtained via~(\ref{eqn-Tp-tbo-formal}) from the matrix elements of $\Toprel$ on the
two-body space.  If $\etavec$ and $\xivec$ represent coordinates or momenta,
then $s$ depends upon the choice of convention in~(\ref{eqn-coords-rcm}), but,
if these variables represent oscillator ladder operators
[see~(\ref{eqn-ladder-rel-cm})], then $s$ is simply unity.

Throughout this work, in spherical tensor coupling notation, we take a vector
operator $\vec{A}$ to indicate the corresponding rank-$1$
covariant spherical tensor operator $A_1$, \textit{e.g.}, $(\etavec\times \xivec)_{L_0}\equiv(\eta_1\times \xi_1)_{L_0}$.  The spherical components of $A_1$ are given
in terms of the Cartesian components, as usual, by~\cite{varshalovich1988:am}
\begin{equation}
\label{eqn-vector-spherical}
A_{1,\pm1}=\mp\frac{1}{\sqrt2}(A_x \pm i A_y)\qquad A_{1,0}=A_z.
\end{equation}

\begin{table*}[t]
  \caption{Physical operators represented in terms of one-body bilinear
    operators $\Top$, together with the relative operator $\Toprel$ entering
    into the two-body representation of the corresponding intrinsic operator
    $\Top'$.  Coefficients and variables are as defined in~(\ref{eqn-O})
    and~(\ref{eqn-Trel-etarel-xirel}).  Only the isoscalar forms of the
    operators are explicitly tabulated, and normalization conventions may vary (see text).  The values for blank entries should be
    taken as identical to those immediately above.}
  \label{tab-physical-bilinear}
\begin{center}
\begin{ruledtabular}
  \begin{tabular}{ll@{~}lll@{~~~~}lll}
    &&\multicolumn{3}{c}{$\Top$}&\multicolumn{3}{c}{$\Toprel$}
    \\
    \cline{3-5}\cline{6-8}
    Operator       & $\kappa$ & $\etavec$ & $\xivec$ & $L_0$ &  $s$ & $\etavecrel$ & $\xivecrel$\\
    \hline
    $\sum_i x_i^2$  & $-\sqrt{3}$ 
    & $\xvec$        &  $\xvec$        & $0$ 
    & $\tBstack{1/2}{1}$ & $\xvecrel$ & $\xvecrel$
    \\
    $Q_2$  & $\sqrt{\frac{15}{8\pi}}$
    &&& $2$
    \\
    $T$  & $-\sqrt{3}\frac{\hbar^2}{2m}$ 
    & $\kvec$        &  $\kvec$        & $0$ 
    & $\tBstack{2}{1}$ & $\kvecrel$ & $\kvecrel$
    \\
    $\Lvec$  & $-i\sqrt{2}$ 
    & $\xvec$        &  $\kvec$        & $1$ 
    & $1$ & $\xvecrel$ & $\kvecrel$
    \\
    $N$ & $-\sqrt{3}$
    & $\cvec^\dagger$ & $\cvec$ & $0$
    & $1$ & $\cvecrel^\dagger$ & $\cvecrel$
    \\
    $\Lvec$~[$=C^{(11)}_1$] & $-\sqrt{2}$
    &&& $1$
    \\
    $\calQ_2$ ~[$=\sqrt{3}C^{(11)}_2$] & $\sqrt{6}$
    &&& $2$
    \\
    $A^{(20)}_0$ & $-\frac{1}{\sqrt{2}}$
    & $\cvec^\dagger$ & $\cvec^\dagger$ & $0$
    & $1$ & $\cvecrel^\dagger$ & $\cvecrel^\dagger$
    \\
    $A^{(20)}_2$ & $+\frac{1}{\sqrt{2}}$
    &&& $2$
    \\
    $B^{(02)}_0$ & $-\frac{1}{\sqrt{2}}$
    & $\cvec$ & $\cvec$ & $0$
    & $1$ & $\cvecrel$ & $\cvecrel$
    \\
    $B^{(02)}_2$ & $+\frac{1}{\sqrt{2}}$
    &&& $2$
    \\
  \end{tabular}
\end{ruledtabular}
\end{center}
\end{table*}


\subsection{Operators of the form $(\xvec\times\xvec)_{L_0}$}
\label{sec-app-physical-x-x}

For the coupled product of a spherical tensor operator with itself, couplings of
odd rank vanish identically by the symmetry properties of the Clebsch-Gordan
coefficients.  Therefore, only the couplings with $L_0=0,2$ arise for
$(\xvec\times\xvec)_{L_0}$.  The isovector (or, rather, proton-only) forms give
rise to the electric monopole ($L_0=0$) and quadrupole ($L_0=2$) operators.

For $L_0=0$, we obtain the squared radius operator, taken already for illustration in Sec.~\ref{sec-background}.  The r.m.s.\
point-nucleon radius relative to the \textit{origin} is obtained as
\begin{math}
  r_m=\bigl[A^{-1}\tbracket{\sum_i x_i^2}\bigr]^{1/2},
\end{math}
where $x_i^2\equiv \xvec_i\cdot\xvec_i$.
For the one-body summed squared radius operator\footnote{While it might be
  tempting to denote the one-body summed squared radius operator by,
  say, $r^2$, this does not provide the basis for a robust notation, given
  likely confusion with the mean summed squared radius operator and the
  r.m.s.\ value itself.}  appearing inside the expectation value, we have
\begin{equation}
  \sum_i x_i^2
  =U\Bigl[(-\sqrt{3})\bigl(\xvec\times\xvec\bigr)_0\Bigr].
\end{equation}
Here we make use of the relation
\begin{math}
  \etavec\cdot\xivec =-\sqrt{3}(\etavec\times\xivec)_0.
\end{math}
between the spherical
tensor product of rank $0$ and the standard vector dot product.
We thus obtain, for the summed squared radius operator, the identifications
$\kappa\rightarrow -\sqrt{3}$, $\etavec\rightarrow\xvec$,
$\xivec\rightarrow\xvec$, and $L_0\rightarrow0$, as given in
Table~\ref{tab-physical-bilinear}.

The r.m.s.\ point-nucleon radius relative to the \textit{center of mass}, as
noted in Sec.~\ref{sec-background}, is obtained by instead using
the corresponding intrinsic coordinate within the expectation value,
\begin{math}
  r'_m=\bigl[A^{-1}\tbracket{\sum_i x_i^{\prime 2}}\bigr]^{1/2}.
\end{math}
The summed squared intrinsic coordinate operator within the expectation value
may then be represented in separable form,
by~(\ref{eqn-Tp-obo-separable-tbo-formal}), as
\begin{equation}
  \begin{aligned}
    \sum_i x_i^{\prime 2}
    &=
      \Bigl(1-\frac{1}{A}\Bigr)
      U\Bigl[
        (-\sqrt{3})\bigl(\xvec\times\xvec\bigr)_{0}
        \Bigr]
      \ifproofpre{\\&\qquad}{}
      -\frac1{A}
      V\Bigl[2
        (-\sqrt{3}) \bigl(\xvec_1 \times \xvec_2\bigr)_{0}
        \Bigr]
      \\
      &=
      \Bigl(1-\frac{1}{A}\Bigr)
      U[x^2]
      -\frac1{A}
      V\bigl[2
        \xvec_1 \cdot \xvec_2
        \bigr],
  \end{aligned}
\end{equation}
where $x^2\equiv \xvec\cdot\xvec$,
or in
manifestly
two-body form, by~(\ref{eqn-Tp-tbo-formal}) and then~(\ref{eqn-coords-rcm}), as
\begin{equation}
  \begin{aligned}
    \sum_i x_i^{\prime 2}
    &=\frac{2}{A}V\Bigl[
      (-\sqrt{3})
      \tfrac12
      \bigl[
        \bigl(\xvec_1-\xvec_2\bigr)
        \times
        \bigl(\xvec_1-\xvec_2\bigr)
        \bigr]_{0}
      \Bigr]
    \\
    &=\frac{2}{A}V\Bigl[
      \tBstack{1/2}{1} x_{\text{rel}}^2
      \Bigr],
  \end{aligned}
\end{equation}
where $x_{\text{rel}}^2\equiv \xvecrel\cdot\xvecrel$.  Similarly, the
expectation value of the squared radius in the intrinsic frame may be
represented in terms of the value calculated using the naive, one-body
laboratory-frame operator, less a center-of-mass contribution,
using~(\ref{eqn-Tp-separated}):
\begin{equation}
  \label{eqn-rsqr-decomposition}
  \Bigl\langle
  \sum_i x_i^{\prime 2}
  \Bigr\rangle
    =
  \Bigl\langle
  \sum_i x_i^{2}
  \Bigr\rangle
  -\Bstack{A}{1}
  \Bigl\langle
  x_{\text{c.m.}}^2
  \Bigr\rangle
  ,
\end{equation}
where $x_{\text{c.m.}}^2\equiv \xveccm\cdot\xveccm$.  This center-of-mass
contribution is trivially known in the case of a factorized $0s$ harmonic
oscillator center-of-mass wave function.

The r.m.s.\ radius of the probability distribution of nucleons of a
single species $\alpha$ (protons or neutrons), relative to the origin, is
instead obtained as (see also
Refs.~\cite{bacca2012:6he-hyperspherical,caprio2014:cshalo})
\begin{equation}
  \label{eqn-ralpha}
  \begin{aligned}
    r_\alpha
    &=\Bigl[\frac{1}{N_\alpha}\tbracket{\sum_i x_i^2 \delta_{i\alpha}}\Bigr]^{1/2}
    \\
    &=\Bigl[\frac{1}{N_\alpha}\tbracket{U\bigl[(-\sqrt{3})(\xvec\times\xvec)_0\delta_\alpha\bigr]}\Bigr]^{1/2}.
  \end{aligned}
\end{equation}
The operator appearing in the expectation value is thus now the proton-only or
neutron-only version of the summed squared radius operator considered above and
listed in Table~\ref{tab-physical-bilinear}, and the results of
Sec.~\ref{sec-bilinear-isovector} now apply.  The corresponding radius relative
to the center of mass (of the nucleus as a whole, that is, not just of the
nucleons of the same species) is obtained as the corresponding intrinsic
operator.  The intrinsic r.m.s.\ radius $r_p$ of the point proton
distribution, in particular, is related, after hadronic physics
corrections~\cite{friar1997:charge-radius-correction}, to the experimentally
accessible nuclear charge radius.  This same proton instrinsic squared radius
operator provides the leading order contribution to the $E0$ transition
operator~\cite{church1956:e0-transitions,bohr1998:v1}.

In general, the intrinsic operator must be used in evaluating observables.
However, for many-body calculations in which the wave function is known to
factorize, with pure harmonic oscillator $0s$ motion for the center of mass, the
naive one-body squared radius or $E0$ operator may be substituted for the
intrinsic two-body operator, \textit{provided} the known contribution arising
from the zero-point motion of the center of mass is subtracted off.  This may be
determined by computing the $0s$ expectation value of the center-of-mass terms
in~(\ref{eqn-Tp-separated}) or~(\ref{eqn-Tivp-foiled-again}), as discussed in
Sec.~\ref{sec-bilinear}.

For $L_0=2$, we obtain the quadrupole operator.  The electromagnetic operator
which induces $E2$ transitions is, in the leading-order or impulse
approximation, the proton-only operator with spherical
components~\cite{eisenberg1976:v3,bohr1998:v1}
\begin{equation}
  \label{eqn-Qp-one-body-Y}
Q_{2,p}=\sum_i e_i x^2_i Y_{2}(\hat{\xvec}_i),
\end{equation}
where $Y_2$ denotes the spherical tensor with the spherical harmonics $Y_{2,\mu}$ as its components, the charge of the $i$th nucleon is given by $e_i=e
\delta_{i,p}$, and $\hat{\xvec}_i$ implicitly represents the polar angles for
the $i$th particle ($\xvec_i=x_i\hat{\xvec}_i$).  For purposes of identifying
the factors in~(\ref{eqn-O}) for Table~\ref{tab-physical-bilinear}, however, we
need only consider the isoscalar (or mass) quadrupole operator, reflecting the combined proton and neutron point-nucleon density,
\begin{equation}
  \label{eqn-Q-one-body-Y}
Q_{2}=\sum_i x^2_i Y_{2}(\hat{\xvec}_i),
\end{equation}
in which we have omitted the electron charge factor $e$ as irrelevant.  Noting
the spherical tensor identity
$x^2Y_2(\hat{\xvec})=[15/(8\pi)]^{1/2}(\xvec\times\xvec)_2$, we thus have
\begin{equation}
  \label{eqn-Q-one-body}
  Q_{2}=U\Big[ \sqrt{\frac{15}{8\pi}} (\xvec\times\xvec)_2\Big],
\end{equation}
giving the identifications $\kappa\rightarrow [15/(8\pi)]^{1/2}$,
$\etavec\rightarrow\xvec$, $\xivec\rightarrow\xvec$, and $L_0\rightarrow2$ in
Table~\ref{tab-physical-bilinear}.\footnote{The operator $Q_{2,p}$ as defined
  in~(\ref{eqn-Qp-one-body-Y}), and thus $Q_2$ as defined
  in~(\ref{eqn-Q-one-body-Y}), is normalized so as to match the $E2$ transition
  operator as it naturally appears in the multipole expansion (\textit{e.g.},
  Refs.~\cite{ring1980-nuclear-many-body,eisenberg1987:v1,suhonen2007:nucleons-nucleus}).
  The static quadrupole \textit{moment} $Q$ is defined via the Cartesian
  quadrupole tensor $Q_{rs}=3x_rx_s-x^2\delta_{rs}$ as the expectation value of
  $\sum_i Q_{zz,i}$ in the stretched state $\tket{JJ}$, giving
  $Q=(16\pi/5)^{1/2}\tme{JJ}{Q_{2,0}}{JJ}$.  Alternatively, the quadrupole
  operator may be normalized (\textit{e.g.}, Ref.~\cite{bohr1998:v1}) as
  $Q^\text{mom}_2= (16\pi/5)^{1/2} x^2
  Y_{2}(\hat{\xvec})=\sqrt{6}(\xvec\times\xvec)_2$, so as to directly give the
  quadrupole moment, yielding instead $\kappa\rightarrow\sqrt{6}$.  The latter
  normalization is consistent with that of the $\grpsu{3}$ quadrupole generator
  $\calQ_2$ defined below [see~(\ref{eqn-Q-su3-sum-bilinear})
    and~(\ref{eqn-Q-su3-sum-spherical-harmonic})].}  The isoscalar and isovector
forms may be combined as usual, by~(\ref{eqn-T-alpha-terms}), to recover
$Q_{2,p}$, which enters into electromagnetic observables, and $Q_{2,n}$, which
enters into nuclear scattering
observables~\cite{bernstein1981:pn-me-hadron-scatt}.

Again, the intrinsic operator must be used in evaluating
observables.  For many-body calculations in which the wave function is known to
factorize, with pure harmonic oscillator $0s$ motion for the center of mass, the
naive one-body quadrupole operator may be used in the evaluation of quadrupole
moments or $E2$ transition matrix elements, since the matrix elements of the
center-of-mass contributions vanish by angular momentum selection rules.  This
property is manifest for the isoscalar operator, from its simple separation
into
intrinsic and center-of-mass quadrupole operators as
\begin{equation}
  Q_2'=Q_2-\Bstack{A}{1}Q_{2,\text{c.m.}},
\end{equation}
by~(\ref{eqn-Tp-separated}).  For the isovector quadrupole operator, the
corresponding property follows from more detailed term-by-term analysis
of~(\ref{eqn-Tivp-foiled-again}), as discussed in
Sec.~\ref{sec-bilinear-isovector}.

\subsection{Operators of the form $(\kvec\times\kvec)_{L_0}$}
\label{sec-app-physical-k-k}

For bilinear operators of the form $(\kvec\times\kvec)_{L_0}$, again only the
couplings with $L_0=0,2$ arise, while the coupling with $L_0=1$ vanishes
identically by the symmetry properties of the Clebsch-Gordan coefficients. The
summed squared momentum operator obtained for $L_0=0$ enters into the familiar
kinetic energy operator $T$, taken already for illustration in
Sec.~\ref{sec-background} in the context of the intrinsic Hamiltonian.  Since
the expressions obtained for this operator closely match those for the summed
squared radius operator above (Sec.~\ref{sec-app-physical-x-x}), it suffices to
quote the results in Table~\ref{tab-physical-bilinear}.

The explicit two-body expression for $T'$, from~(\ref{eqn-Tp-tbo-formal}), is
\begin{equation}
T'=
\frac{2}{A}
\,
V \Bigl[
\underbrace{
  \frac12
  \frac{\hbar^2}{2m}
\bigl[
 \bigl(\kvec_1-\kvec_2\bigr)
\cdot
 \bigl(\kvec_1-\kvec_2\bigr)
\bigr]
}_{\Trel}
\Bigr],
\end{equation}
where, in terms of the relative coordinate,
\begin{equation}
\label{eqn-Trel-kinetic}  
\Trel=
  \frac{\hbar^2}{2m}
  \Bstack{2}{1}
  \kvecrel\cdot\kvecrel.
\end{equation}
Then, we note the one-body plus
separable two-body form of the intrinsic kinetic energy obtained from~(\ref{eqn-Tp-obo-separable-tbo-formal}),\footnote{The comparative clarity of notation afforded
by the one-body and two-body operator conventions of Sec.~\ref{sec-nbody} may be noted by
comparison of~(\ref{eqn-Tp-obo-separable-tbo-formal-kinetic}) to prior expressions (see, \textit{e.g.}, Appendix~A of
Ref.~\cite{caprio2012:csbasis}).}
\begin{equation}
\label{eqn-Tp-obo-separable-tbo-formal-kinetic}  
T'=
\frac{\hbar^2}{2m}
\Bigl(1-\frac{1}{A}\Bigr)
U\bigl[
 k^2
\bigr]
-
  \frac{\hbar^2}{2m}
\frac1{A}
V\bigl[
 2\kvec_1 \cdot \kvec_2
\bigr],
\end{equation}
which may be used to evaluate the two-body matrix elements of $T'$ in terms of
one-body matrix elements of $\kvec$ and $k^2$.

Incidentally, the alternative prefactors in braces in~(\ref{eqn-Trel-kinetic}),
arising as the factor $s$ from~(\ref{eqn-Trel-etarel-xirel}), may be interpreted
as introducing the reduced mass in the relative kinetic energy, as can be seen
when the preceding expression is rearranged as
\begin{equation}
\label{eqn-Trel-kinetic-reduced-mass}  
\Trel=
\frac{\hbar^2}{2\lbrace m/2,m\rbrace}
  \kvecrel\cdot\kvecrel.
\end{equation}
The reduced mass is thus $m/2$ under the mechanics convention for the relative
coordinate but is simply the ordinary mass under the symmetric convention~---
this distinction in turn relates to different values for the relative oscillator
length $\brel$ arising below in~(\ref{eqn-raising-cm}) under these two conventions.

\subsection{Operators of the form $(\xvec\times\kvec)_{L_0}$}
\label{sec-app-physical-x-k}

All couplings $L_0=0,1,2$ are possible for $(\xvec\times\kvec)_{L_0}$ or
$(\kvec\times\xvec)_{L_0}$.  These two operators $(\xvec\times\kvec)_{L_0}$
and $(\kvec\times\xvec)_{L_0}$ are equivalent, to within a possible phase
factor, by the spherical tensor coupled
commutator relations~(\ref{eqn-commutator-coupled}) and~(\ref{eqn-commutator-coupled-canonical}), except in the scalar case
$L_0=0$, where they differ by a nonzero constant ($c$-number) commutator:
\begin{math}
  (\xvec\times\kvec)_{L_0}
  =
  (-)^{L_0}(\kvec\times\xvec)_{L_0}
  -\sqrt{3}i\delta_{L_0,0}.
\end{math}

The
coupling with $L_0=1$, in its isoscalar form, yields the orbital angular
momentum operator $\Lvec$.
That is, the total orbital angular momentum of the nucleons, relative to the origin, is represented by the operator
\begin{equation}
  \label{eqn-L-def-coord}
  \Lvec
  =\sum_i \xvec_i\times \kvec_i
  =U\bigl[(-i\sqrt{2})(\xvec\times\kvec)_1\bigr],
\end{equation}
which we may equivalently denote in spherical tensor form as the rank-$1$ spherical tensor $L_1$.
Here we make use of the relation
\begin{math}
  \etavec\times\xivec =-i\sqrt{2}(\etavec\times\xivec)_1
\end{math}
between the spherical tensor product of rank $1$ and the standard vector cross
product.  Note that we consider here the dimensionless angular momentum
operator, with dimensionless eigenvalues $L(L+1)$, commonly encountered in the nuclear
many-body literature, in terms which the usual physical angular momentum operator may
be recovered as $\Lvec^{\text{phys}}=\hbar\Lvec$.  We thus obtain the
identifications $\kappa\rightarrow -i\sqrt{2}$, $\etavec\rightarrow\xvec$,
$\xivec\rightarrow\kvec$, and $L_0\rightarrow1$, as given in
Table~\ref{tab-physical-bilinear}.

The isoscalar and isovector forms may be combined as usual, by~(\ref{eqn-T-alpha-terms}), to extract the proton and
neutron orbital angular momentum operators separately.  These enter, along with
the proton and neutron spin operators, into the leading-order or impulse-approximation $M1$ transition
operator as~\cite{eisenberg1976:v3,bohr1998:v1,rowe2010:collective-motion}
\begin{equation}
  \label{eqn-M1}
  \vec{M}_1=\sqrt{\frac{3}{4\pi}} \mu_N
  \bigl(
    \glp \Lvec_{p}+\gln \Lvec_{n}
    +\gsp \Svec_{p}+\gsn \Svec_{n}
    \bigr),
\end{equation}
where $\glp=1$, $\gln=0$, $\gsp\approx 5.586$, and $\gsn \approx-3.826$.

As usual, the appropriate operator for evaluating the physical $M1$ transition
matrix element is the corresponding intrinsic operator
$\vec{M}'_1$~\cite{eisenberg1976:v3}, which is defined in terms of the two-body
intrinsic angular momenta $\Lvec'_{p}$ and $\Lvec'_{n}$.  Two-body matrix
elements of these operators for use in a many-body calculation may thus be
evaluated either from relative two-body matrix elements or by the separable
approach, again using generic results for bilinear operators
(Sec.~\ref{sec-bilinear}).

The one-body spin operators $\Svec_{p}$ and $\Svec_{n}$ are already
Galilean-invariant, so their contributions to the $\vec{M}_1$ operator are
already those found in the c.m.\ frame.  However, for computational convenience,
these operators may also be reexpressed as $A$-dependent two-body operators (the
requisite expressions are provided for reference in
Appendix~\ref{sec-app-spin}), so that $\vec{M}'_1$ may be represented as a pure
two-body operator and thus entirely expressed in terms of two-body matrix
elements.

When the wave function is known to factorize with pure harmonic oscillator $0s$
motion for the center of mass, the naive one-body form of the $M1$ operator
given in~(\ref{eqn-M1}) may be used.  In this case, the matrix elements of the
center-of-mass contributions vanish by angular momentum selection rules, by the
analysis of Sec.~\ref{sec-bilinear-isovector}, much as noted for the quadrupole
operator above (Sec.~\ref{sec-app-physical-x-x}).

\subsection{Bilinears in harmonic oscillator ladder operators}
\label{sec-app-physical-ladder}

The bilinear couplings $(\cvec^\dagger\times\cvec)_{L_0}$ or
$(\cvec\times\cvec^\dagger)_{L_0}$ of an oscillator creation and
annihilation\footnote{In these spherical tensor coupled products, the symbol $\cvec$ represents the
  rank-$1$ covariant spherical tensor operator $c_1$ with components
  obtained from the Cartesian vector $\cvec$ (\textit{e.g.}, Sec.~5.8 of
  Ref.~\cite{rowe2010:rowanwood}) by~(\ref{eqn-vector-spherical}).  Care must be taken in comparison with the
  literature, where the
  spherical tensor $\tilde{c}_1$, defined as the covariant adjoint of
  $c_1^\dagger$, is commonly used, yielding expressions of the form, \textit{e.g.},
  $(c_1^\dagger\times\tilde{c}_1)_{L_0}$.
  We may have either $\tilde{c}_1=\pm c_1$, depending upon the convention adopted for the
  covariant adjoint in a given reference, as detailed
  in Sec.~\ref{sec-app-ladder-defn}.} operator are possible for $L_0=0,1,2$, while
$(\cvec^\dagger\times\cvec^\dagger)_{L_0}$ and $(\cvec\times\cvec)_{L_0}$ are
only nonvanishing for $L_0=0,2$.  The expressions
$(\cvec^\dagger\times\cvec)_{L_0}$ and $(\cvec\times\cvec^\dagger)_{L_0}$ are
equivalent, to within a phase factor, by the spherical tensor coupled
commutator relations~(\ref{eqn-commutator-coupled})
and~(\ref{eqn-commutator-coupled-canonical}), except in the scalar case $L_0=0$,
where they differ by a $c$-number commutator:
\begin{math}
  (\cvec\times\cvec^\dagger)_{L_0}
  =
  (-)^{L_0}(\cvec^\dagger\times\cvec)_{L_0}
  -\sqrt{3}\delta_{L_0,0}.
\end{math}
The definitions and properties of the single-particle, relative,
and intrinsic forms of the harmonic oscillator ladder operators are reviewed in
Appendix~\ref{sec-app-ladder}.  Bilinears in the coordinates and/or momenta may, of
course, be reexpressed in terms of bilinears in oscillator ladder operators via
the relations~(\ref{eqn-ladder-inverse-b}), giving, \textit{e.g.},
\begin{equation}
  b^{-2} Q_2
  =
  \sqrt{\frac{15}{8\pi}}
  \bigl[
  (\cvec^\dagger\times\cvec)_2
  +\tfrac12(\cvec^\dagger\times\cvec^\dagger)_2 +
  \tfrac12(\cvec\times\cvec)_2
  \bigr]
\end{equation}
for the mass quadrupole operator of Sec.~\ref{sec-app-physical-x-x}.

Generators for the $\grpu{3}$ group of a three-dimensional harmonic oscillator
are given in Cartesian form by $C_{rs}=\tfrac12( c_r^\dagger c_s + c_s
c_r^\dagger) = c_r^\dagger c_s + \tfrac12 \delta_{rs}$ ($r,s=x,y,z$).  Elliot's
realization of $\grpu{3}$ for the nuclear
problem~\cite{elliott1958:su3-part1,*elliott1958:su3-part2,*elliott1963:su3-part3,*elliott1968:su3-part4,harvey1968:su3-shell}
makes use of the physical subgroup chain
$\grpu{3}=\grpu{1}\times[\grpsu{3}\supset \grpso{3}]$, which incorporates the
$\grpso{3}$ orbital angular momentum group.  The set of $9$ Cartesian generators
$C_{rs}$ is then more conveniently transformed to a set of spherical tensor
generators,
\begin{equation}
  \label{eqn-su3-generators}
  \begin{aligned}
    H_0&= \tfrac12(\cvec^\dagger\cdot\cvec+\cvec\cdot\cvec^\dagger)=N+\tfrac32\\
    L_1&=-\sqrt{2}(\cvec^\dagger\times\cvec)_1\\
    \calQ_2&=\sqrt{6}(\cvec^\dagger\times\cvec)_2,
  \end{aligned}
\end{equation}
again with a total of nine components.  Here the (dimensionless) harmonic
oscillator Hamiltonian $H_0$ is the generator of the trivial Abelian $\grpu{1}$
group, $L_1$ is the familiar orbital angular momentum operator
of~(\ref{eqn-L-def-coord}), now written in terms of ladder operators, and
$\calQ_2$ is the $\grpsu{3}$ quadrupole tensor.  These operators close under
commutation, with spherical tensor coupled commutators
$[L_1,L_1]_1=-\sqrt{2}L_1$, $[\calQ_2,L_1]_2=-\sqrt{6}\calQ_2$, and
$[\calQ_2,\calQ_2]_1=3\sqrt{10}L_1$, and all other commutators
vanishing~\cite{harvey1968:su3-shell}.  The resulting
bilinear operators of the type $(\cvec^\dagger\times\cvec)_{L_0}$ are summarized
in Table~\ref{tab-physical-bilinear}.  For the $A$-body system, the Elliott
$\grpu{3}$ generators are realized as the corresponding one-body operators, that
is, summed over nucleons.

In particular, for $L_0=0$, the $A$-particle harmonic oscillator Hamiltonian,
\begin{equation}
    H_0
    =\sum_i \tfrac12\bigl(\cvec_i^\dagger\cdot\cvec_i+\cvec_i\cdot\cvec_i^\dagger\bigr)
    =N+\tfrac32A,
\end{equation}
is related to the $A$-particle harmonic oscillator number operator $N$ by a
$c$-number offset $\tfrac32A$, representing the aggregate zero-point energies of
the $A$ particles.  The number operator is simply
\begin{equation}
  N
  =\sum_i \cvec_i^\dagger\cdot\cvec_i
  \\
  =U\Bigl[(-\sqrt{3})\bigl(\cvec^\dagger\times\cvec\bigr)_0\Bigr].
\end{equation}
We thus obtain the identifications
$\kappa\rightarrow -\sqrt{3}$, $\etavec\rightarrow\cvec^\dagger$,
$\xivec\rightarrow\cvec$, and $L_0\rightarrow0$, as given in
Table~\ref{tab-physical-bilinear}.

For $L_0=2$, the $\grpsu{3}$ quadrupole operator may be reexpressed in terms of
the coordinate and momentum bilinears of
Secs.~\ref{sec-app-physical-x-x}--\ref{sec-app-physical-k-k}, which are then scaled by appropriate
powers of the oscillator length $b$ from~(\ref{eqn-ho-hamiltonian-b}) to produce
a dimensionless result, as
\begin{equation}
  \label{eqn-Q-su3-sum-bilinear}
    \calQ_2=\sum_i \sqrt{\tfrac32}\bigl[b^{-2}(\xvec_i\times\xvec_i)_2+b^{2}(\kvec_i\times\kvec_i)_2],
\end{equation}
and thus~\cite{harvey1968:su3-shell}
\begin{equation}
  \label{eqn-Q-su3-sum-spherical-harmonic}
  \calQ_2=\sum_i \sqrt{\frac{4\pi}{5}}
  \bigl[
    b^{-2} x^2_i Y_{2}(\hat{\xvec}_i)
    +
    b^{2} k^2_i Y_{2}(\hat{\kvec}_i)
    \bigr].
\end{equation}
The $\grpsu{3}$ quadrupole tensor may thus be recognized as a linear combination
of the mass quadrupole operator of~(\ref{eqn-Q-one-body-Y}) and its momentum-space
analog.

An important consequence of the bilinear forms of these $\grpu{3}$ generators is
that, by~(\ref{eqn-Tp-separated}), each generator separates into intrinsic and
center-of-mass parts~\cite{kretzschmar1960:su3-shell-part2-com}.  That is,
rearranging~(\ref{eqn-Tp-separated}) to isolate the one-body operator on the
left-hand side, we have
\begin{maybemultline}
  \label{eqn-Tp-separated-rearranged}
  \underbrace{\sum_i \bigl(\etavec_i\times\xivec_i\bigr)_{L_0}}_{\Top}
  =
  \underbrace{\sum_i \bigl(\etavec'_i\times\xivec'_i\bigr)_{L_0}}_{\Top'}
  \ifproofpre{\\}{}
  +
  \underbrace{
    \frac1A  \Bigl[\bigl(\sum_{i}\etavec_i\bigr) \times \bigl(
      \sum_{j}\xivec_j\bigr)\Bigr]_{L_0}
  }_{\Topcm},
\end{maybemultline}
yielding a decomposition of a general bilinear operator $\Top$ into intrinsic
and c.m.\ parts as $\Top=\Top'+\Topcm$.  Thus, in particular,
$H_0=H_0'+H_{0,\text{c.m.}}$ (and, similarly, $N=N'+N_{\text{c.m.}}$),
$\Lvec=\Lvec'+\Lvec_{\text{c.m.}}$, and
$\calQ_2=\calQ_2'+\calQ_{\text{c.m.},2}$.  When the separation is made into
intrinsic and center-of-mass parts of the $\grpu{1}$ generator $H_0$, then
$H_{0,\text{c.m.}}=\Ncm+3/2$ for the center-of-mass oscillator, while the
remaining $3(A-1)/2$ units of zero-point energy reside in the intrinsic
Hamiltonian, which is related to the intrinsic number operator by
$H_0'=N'+3(A-1)/2$.

The $\grpu{3}$ symmetry group of the harmonic oscillator lies within the larger
$\grpsptr$ dynamical group of the oscillator, which provides raising and
lowering operators connecting different oscillator shells
(\textit{e.g.}, Ref.~\cite{wybourne1974:groups}).  In Cartesian form, the
generators of $\grpsptr$ consist of the symplectic raising operators
$A_{rs}=A_{sr}=c^\dagger_r c^\dagger_s$, which carry $+2$ oscillator quanta, and
the symplectic lowering operators $B_{rs}=B_{sr}=c_r c_s$, which carry $-2$
oscillator quanta, along with the number
conserving $\grpu{3}$ generators $C_{rs}$ defined above, giving a total of
twenty-one components.

To make use of the full machinery of tensor operators in a symmetry adapted
many-body basis, it is more useful to transform these generators to
$\grpsu{3}\supset\grpso{3}$-coupled form.  This is accomplished making use of
the property that the oscillator creation and annihilation operators form
$(1,0)$ and $(0,1)$ $\grpsu{3}$ tensors, respectively.  Written as $\grpsu{3}$
coupled products, the generators are
then~\cite{rosensteel1992:sp3r-tensors-gtssnp91,escher2002:pds-symplectic}\footnote{Here
  $T^{(\lambda,\mu)}_L$ represents the spherical tensor component with
  $\grpso{3}$ angular momentum $L$ of an $\grpsu{3}$ tensor forming an irrep
  with Elliott labels $(\lambda,\mu)$~\cite{wybourne1974:groups}.}
\begin{equation}
  \label{eqn-sp3r-generators-su3-coupled}
  \begin{gathered}
  \begin{aligned}
    A^{(2,0)}_{L}&=\frac{1}{\sqrt{2}} \bigl( \cdsu \times \cdsu \bigr)^{(2,0)}_L
    &\quad L&=0,2
    \\
    B^{(0,2)}_{L}&=\frac{1}{\sqrt{2}} \bigl( \csu \times \csu \bigr)^{(0,2)}_L
    &\quad L&=0,2
    \\
    C^{(1,1)}_{L}&=\sqrt{2} \bigl( \cdsu \times \csu \bigr)^{(1,1)}_L
    &\quad L&=1,2
  \end{aligned}
  \\
    H^{(0,0)}_{0}=
      \frac{\sqrt{3}}{2}\Bigl[
        \bigl( \cdsu \times \csu \bigr)^{(0,0)}_0
        \ifproofpre{\\\qquad\qquad}{}
        +
        \bigl( \csu \times \cdsu \bigr)^{(0,0)}_0
        \Bigr].
  \end{gathered}
\end{equation}
Using $\grpsu{3}\supset\grpso{3}$ reduced coupling coefficients under standard
phase conventions~\cite{draayer1973:su3-cg} to explicitly evaluate the
$\grpsu{3}$ coupled products in~(\ref{eqn-sp3r-generators-su3-coupled}) gives
\begin{equation}
  \label{eqn-sp3r-generators-branched}
  \begin{gathered}
  \begin{aligned}
    A^{(2,0)}_{0}&=-\frac{1}{\sqrt{2}} \bigl( \cvec^\dagger \times \cvec^\dagger \bigr)_0
    &\qquad
    A^{(2,0)}_{2}&=+\frac{1}{\sqrt{2}} \bigl( \cvec^\dagger \times \cvec^\dagger \bigr)_2
    \\
    B^{(0,2)}_{0}&=-\frac{1}{\sqrt{2}} \bigl( \cvec \times \cvec \bigr)_0
    &
    B^{(0,2)}_{2}&=+\frac{1}{\sqrt{2}} \bigl( \cvec \times \cvec \bigr)_2
    \\
    C^{(1,1)}_{1}&=-\sqrt{2} \bigl( \cvec^\dagger \times \cvec \bigr)_1
    &
    C^{(1,1)}_{2}&=+\sqrt{2} \bigl( \cvec^\dagger \times \cvec \bigr)_2
  \end{aligned}
  \\
    H^{(0,0)}_{0}=
      -\frac{\sqrt{3}}{2}\Bigl[
        \bigl( \cvec^\dagger \times \cvec \bigr)_0
        +
        \bigl( \cvec \times \cvec^\dagger \bigr)_0
        \Bigr],
  \end{gathered}
\end{equation}
from which we recognize $C^{(1,1)}_{1}=L_1$, $\sqrt{3}C^{(1,1)}_{2}=\calQ_2$,
and $H^{(0,0)}_{0}=H_0=N+3/2$.  The correspondence to the generic bilinear
operator is summarized in Table~\ref{tab-physical-bilinear}.  Again, as for the
$\grpu{3}$ generators above, the generators acting on an $A$-body system are
obtained as the resulting one-body operators, summed over nucleons.
The intrinsic forms of the $\grpsptr$ generators given in~(38) of
Ref.~\cite{rosensteel1980:sp6r-shell} or~(2) of
Ref.~\cite{escher1998:sp6r-shell-su3coupling} may be recognized as the
intermediate step of our decomposition found in~(\ref{eqn-Tp-foiled-again}).


\section{Proton-neutron mass difference and corrections to kinetic energy}
\label{sec-app-isovector-kinetic}

Throughout this work, we have assumed that the proton and neutron masses could
be approximated as a single ``nucleon mass'' $m_N$ (see
footnote~\ref{footnote-pn-mass}).
It is well known that including the proton-neutron mass difference induces an
isovector correction to the kinetic energy~\cite{henley1969:isospin-nuclear-forces}.
Here we derive that correction more completely and extend the derivation to the intrinsic
kinetic energy.

Let us define the mean nucleon mass and the nucleon mass deviation by
\begin{equation}
    m_N = \frac{1}{2} (m_p + m_N) \qquad \Delta_m = \frac{1}{2} (m_p - m_n).
\end{equation}
We can then define the relative mass deviation as
\begin{equation}
  \delta_m
  = \frac{\Delta_m}{m_N}
  = \frac{m_p - m_n}{m_p + m_n},
\end{equation}
and write the mass of the $i$th nucleon as
\begin{equation}
  \label{eqn-mass-tau}
  m_i
  = m_N \qty(1 + \tau_{i0} \delta_m).
\end{equation}
The total nuclear mass is then
\begin{equation}
  M=[A + (Z-N)\delta_m]m_N.
\end{equation}

We now examine the total kinetic energy of a system of $A$ nucleons, defined by
\begin{equation}
    T=U \qty[\frac{\hbar^2 k^2}{2m}]=\sum_i \frac{\hbar^2 k_i^2}{2m_i}.
\end{equation}
Expressing the nucleon-dependence of the mass in terms of $\tau_{i0}$ via~(\ref{eqn-mass-tau}), this becomes
\begin{equation}
    T=\sum_i \frac{\hbar^2 k_i^2}{2m_N \qty(1 + \tau_{i0} \delta_m)}.
\end{equation}
Expanding the denominator of the summand as a geometric series, and resumming
the series,\footnote{
  Namely, recognizing that $(\tau_{i0})^2=1$, we obtain
  \begin{math}
    (1+\tau_{i0} \delta_m)^{-1}
        = \sum_{k=0}^\infty (-\tau_{i0})^k \delta_m^k
        = \sum_{\ell=0}^\infty ((-\tau_{i0})^{2\ell} \delta_m^{2\ell} + (-\tau_{i0})^{2\ell+1} \delta_m^{2\ell+1})
        = \sum_{\ell=0}^\infty (1-\tau_{i0}\delta_m)(\delta_m^2)^{\ell}
        = \sfrac{(1-\tau_{i0} \delta_m)}{(1-\delta_m^2)}.
  \end{math}
}
removes $\tau_{i0}$-dependence from the denominator, giving
\begin{equation}
    \label{eqn-total-kinetic-pn}
    T
      = \frac{1}{1-\delta_m^2} \underbrace{\sum_i \frac{\hbar^2 k_i^2}{2 m_N}}_{T_\text{IS}}
            - \frac{\delta_m}{1-\delta_m^2} \underbrace{\sum_i \frac{\hbar^2 k_i^2\tau_{i0}}{2 m_N}}_{T_\text{IV}}.
\end{equation}
We thus obtain an expression for the kinetic energy which separates into
isoscalar and isovector terms.  The prefactor $(1-\delta_m^2)^{-1}$ is
second-order in the nucleon mass difference and is thus neglected in~(2.18)
of Ref.~\cite{henley1969:isospin-nuclear-forces}.

Now we turn to the intrinsic kinetic energy $T'$, obtained by the substitution
$\kvec_i \rightarrow \kvec'_i$.  However, recall that the form of the intrinsic
coordinates and momenta which we have thus far been using in such substitutions,
given in~(\ref{eqn-coords-intrinsic-explicit-sum}), is obtained by neglecting
the proton-neutron mass difference.  We must instead return to the fundamental
definition, given in~(\ref{eqn-coords-intrinsic-mass}), which may be reexpressed
in terms of the masses $m_i$ as
\begin{equation}
    \label{eqn-coords-intrinsic-mass-explicit-sum}
    \xvec'_i=\xvec_i-        {\frac{1}{M}}  \sum_j m_j \xvec_j
    \qquad
    \kvec'_i=\kvec_i-        {\frac{m_i}{M}}  \sum_j \kvec_j.
\end{equation}
Thus, we obtain an intrinsic kinetic energy
\begin{equation}
  T'
  = \sum_i \frac{\hbar^2}{2m_i}\qty(\kvec_i - \frac{m_i}{M}\sum_j \kvec_j)\cdot\qty(\kvec_i - \frac{m_i}{M}\sum_k \kvec_k).
\end{equation}
Expanding, and using the identity $\sum_i m_i = M$ and to recognize and combine like terms, we arrive at
\begin{equation}
  \label{eqn-Tp-separated-kinetic-pn}
    T' = \sum_i \frac{\hbar^2 k_i^2}{2m_i} - \frac{\hbar^2}{2M}\qty(\sum_i \kvec_i)\cdot\qty(\sum_j \kvec_j).
\end{equation}
This expression is recognizable as a decomposition of the intrinsic operator
as the total (one-body) operator less a center-of-mass contribution, generalizing~(\ref{eqn-Tp-separated-kinetic}) to unequal nucleon masses.

Then, to obtain a separation into one-body and separable two-body parts, we
multiply out the product of sums appearing in the second term
of~(\ref{eqn-Tp-separated-kinetic-pn}) and extract the diagonal terms
from the resulting double sum, while eliminating the explicit nucleon mass
dependence from the denominator of the first term as in~(\ref{eqn-total-kinetic-pn}), to obtain
\begin{multline}
  T' = \qty(\frac{1}{1-\delta_m^2}-\frac{1}{A+(Z-N)\delta_m})
  \underbrace{
    \sum_i \frac{\hbar^2 k_i^2}{2 m_N}
  }_{T_\text{IS}}
  \\
  - \frac{\delta_m}{1-\delta_m^2}
  \underbrace{
    \sum_i \frac{\hbar^2 k_i^2\tau_{i0}}{2 m_N}
  }_{T_\text{IV}}
  \ifproofpre{\\}{}
  - \frac{1}{A+(Z-N)\delta_m} \sumprime_{ij} \frac{\hbar^2 \kvec_i \cdot \kvec_j}{2m_N},
\end{multline}
thereby generalizing~(\ref{eqn-Tp-obo-separable-tbo-sum-kinetic}) to unequal
nucleon masses.  In the one-body part, we recognize a separation into isoscalar and isovector terms, involving the same operators $T_\text{IS}$ and $T_\text{IV}$, respectively,
as in~(\ref{eqn-total-kinetic-pn}).  The two-body part
is isoscalar and simply reflects the nucleon mass difference through its
$(Z-N)$-dependent prefactor.  However, if the nucleon mass were instead taken as $m_N
=(Zm_p + Nm_n)/A$ (see footnote~\ref{footnote-pn-mass}), note that the
prefactor of the two-body term would then simply be $1/A$, and the two-body
contribution would reduce to that in~(\ref{eqn-total-kinetic-pn}).


\section{Summation identities for two-body operators}
\label{sec-app-sum}

A couple of straightforward summation identities are essential in working with
one-body and two-body operators.

In particular, any one-body operator may be ``upgraded'' to a
two-body operator.  Observe that
\begin{equation}
\label{eqn-upgrade-sum}
\sumprime_{ij}(u_i+u_j)=2(A-1)\sum_iu_i,
\end{equation}
and thus we recognize that
\begin{equation}
\label{eqn-upgrade}
V[u_1+u_2]=(A-1)U[u].
\end{equation}

In the definition~(\ref{eqn-tbo}) of the two-body operator, the expression
$v_{ij}$ is only taken to be defined for $i\neq j$.  However, in the applications
considered here, where $v_{ij}$ arises as a product operator of the form
$v_{ij}=\eta_i \xi_j$, we also naturally encounter unrestricted sums of $v_{ij}$
over the two particle indices $i$ and $j$.  Such a sum can be broken into
one-body and two-body parts, as
\begin{equation}
\label{eqn-diagonal-sum}
\sum_{ij}v_{ij}=\sum_i v_{ii}+\sumprime_{ij}v_{ij},
\end{equation}
by extracting the diagonal ($i=j$) terms from the double sum.  If $v_{ij}$ is
known to be symmetric under interchange of the particle indices
($i\leftrightarrow j$), the expression on the right hand side can immediately be
recognized as $U[u]+2V[v]$, where $u$ is defined on the single-particle space as
$u=v_{11}$, and $v$ is defined as usual on the two-particle space as
$v=v_{12}=v_{21}$.  More generally, if $v_{ij}$ cannot be assumed to be
symmetric, we must symmetrize the summand of the double sum, to obtain
\begin{equation}
\label{eqn-diagonal-sum-symm}
\sum_{ij}v_{ij}=\sum_i v_{ii}+\tfrac12\sumprime_{ij}(v_{ij}+v_{ji}),
\end{equation}
and recognize the sums on the right hand side as representing one-body and
two-body operators, respectively, to obtain
\begin{equation}
\label{eqn-diagonal-formal}
\sum_{ij}v_{ij}=U[v_{11}]+V[v_{12}+v_{21}].
\end{equation}

\section{Two-body realization of spin operators}
\label{sec-app-spin}

The total spin operators (proton and neutron spin or, equivalently, isoscalar
and isovector spin) are inherently independent of the c.m.\ coordinate degree of
freedom.  They are thus already intrinsic operators.  Nonetheless, they are taken
in linear combination with the orbital angular momentum operators to generate
the $M1$ operator, and, while the intrinsic orbital angular momentum operators
are two-body operators, these spin operators are only one-body operators.  To
incorporate them into the evaluation of two-body matrix elements for the $M1$
operator, it is convenient to use the relation~(\ref{eqn-upgrade}) to upgrade
them to two-body operators.

Thus, we note that the one-body total spin operator
\begin{equation}
\label{eqn-S-obo}
\Svec=U[\svec]=\sum_i\svec_i
\end{equation}
may be rewritten, using identity~(\ref{eqn-upgrade-sum}), as an $A$-dependent
two-body operator, as
\begin{equation}
\label{eqn-S-tbo}
\Svec
=\frac{1}{A-1}\tfrac12\sumprime_{ij} (\svec_i+\svec_j)
=\frac{1}{A-1}V[\Svecrel],
\end{equation}
where $\Svecrel=\svec_1+\svec_2$ is the total spin operator on the two-body
system.

Then, the one-body proton and neutron spin operators are
\begin{equation}
\label{eqn-S-alpha-obo}
\Svec_\alpha=U[\svec\delta_\alpha]=\sum_i\svec_i\delta_{i,\alpha}.
\end{equation}
These operators separate into manifestly isoscalar and isovector parts, much as
for the proton or neutron dipole operator in~(\ref{eqn-D-alpha-terms}).
Inserting the definition~(\ref{eqn-delta-alpha}) for $\delta_\alpha$ gives
\begin{equation}
\label{eqn-S-alpha-terms}
\Svec_\alpha=
\tfrac12
\underbrace{\biggl[\sum_i \svec_i \biggr]}_{\Svec}
+\tfrac12\alpha
\underbrace{\biggl[\sum_i \svec_i \tau_{i0}\biggr]}_{\Sveciv},
\end{equation}
that is, $\Svec_\alpha=\tfrac12\Svec+\tfrac12\alpha\Sveciv$, where $\Svec$ is
simply the isoscalar total spin operator from~(\ref{eqn-S-obo}), while
\begin{equation}
\label{eqn-Siv-obo}
\Sveciv=U[\svec\tau_0]=\sum_i \svec_i \tau_{i0}
\end{equation}
is the isovector spin operator.  Again using identity~(\ref{eqn-upgrade-sum}),
the isovector spin operator may be rewritten as an $A$-dependent two-body
operator, as
\begin{equation}
\label{eqn-Siv-tbo-sum}
\Sveciv
=\frac{1}{A-1}\tfrac12\sumprime_{ij} (\svec_i \tau_{i0}+\svec_j\tau_{j0})
=\frac{1}{A-1}V[\Svecivrel],
\end{equation}
where $\Svecivrel=\svec_1\tau_{01}+\svec_2\tau_{02}$ is the isovector spin
operator on the two-body system.


\section{Isospin-reduced matrix elements: Dipole and bilinear operators}
\label{sec-app-isospin}

Since the spatial and isospin dependences of the isovector parts of the relative
dipole operator (Sec.~\ref{sec-dipole}) and relative intrinsic bilinear operator
(Sec.~\ref{sec-bilinear}) factorize, it is straightforward to evaluate the
isospin contribution to two-body matrix elements.  Let us write the two-body
states as $\tket{\gamma T}$, with isospin $T=0,1$, where $\gamma$ represents
all quantum numbers other than isospin (typically, for an $LS$-coupled relative
oscillator basis, $\gamma\equiv\Nrel LSJ$).  Since we are working with basis
states of good isospin and isovector operators (\textit{i.e.}, spherical tensors
of rank $1$ under isospin rotations, carrying definite isospin $T_0=1$) it is
most streamlined to work with isospin-reduced matrix elements
$\trme{\gamma'T'}{\cdots}{\gamma T}$.

The reduced matrix element $\trme{\gamma'T'}{\Dvecivrel}{\gamma T}$, of the
isovector relative dipole operator $\Dvecivrel$, factorizes into a spatial-spin
part $\propto\tme{\gamma'}{\xvecrel}{\gamma}$, which depends on the particular
choice of basis wave functions, and an isospin part $\trme{T'}{\tauvec_{1} -
  \tauvec_{2}}{T}$, which is independent of the details of the basis.  We can
thus note for reference these latter matrix elements on the two-body system.
Since the operator $\tauvec_{1} - \tauvec_{2}$ is isovector, the matrix element
between $T=0$ states vanishes by the angular momentum triangularity selection
rule:
\begin{math}
\trme{0}{\tauvec_{1} - \tauvec_{2}}{0}=0.
\end{math}
The operator likewise has vanishing expectation value within a $T=1$ state:
\begin{math}
\trme{1}{\tauvec_{1} - \tauvec_{2}}{1}=0.
\end{math}
(The vectorial interpretation is that this state represents the aligned coupling
of the two $T=1/2$ isospins of the nucleons.  The isospin projections of the two
nucleons along any axis in isospin space must therefore align, and their
difference along the axis thus cancels.)  The remaining matrix element, between
$T=0$ and $T=1$ states, may be evaluated by standard angular momentum coupling
methods (namely, Racah's reduction formulas), giving
\begin{math}
\trme{0}{\tauvec_{1} - \tauvec_{2}}{1}=-2\sqrt{3}.
\end{math}
We follow the normalization and phase convention of
Rose~\cite{rose1957:am} for the Wigner-Eckart theorem when defining the reduced
matrix element, \textit{i.e.},
\begin{math}
\tme{J'M'}{T_{J_0M_0}}{JM}=\tcg{J}{M}{J_0}{M_0}{J'}{M'}\trme{J'}{T_{J_0}}{J}.
\end{math}
To convert isospin-reduced matrix elements to the normalization convention of
Edmonds~\cite{edmonds1960:am}, they may be multipled by $(2T'+1)^{1/2}$,
\textit{i.e.}, in this case just unity.

The situation is, in fact, much simpler for the bilinear opertors.  Recall,
from Sec.~\ref{sec-bilinear-isovector}, that we must evaluate two-body matrix
elements of the isoscalar relative operator $\Toprel$~(\ref{eqn-Trel}) and its
isovector counterpart $\Toprel\,(2T_0)$, in the two-body basis.  The
isospin-reduced matrix elements of these two operators are closely related:
\begin{equation}
\label{eqn-rme-Tivrel}
\trme{\gamma'T'}{\Toprel\,(2\vec{T})}{\gamma T}
=\tme{\gamma'}{\Toprel}{\gamma}
\trme{T'}{2\vec{T}}{T}.
\end{equation}
The isospin-reduced matrix element follows from the well-known identity for the
reduced matrix element of the angular momentum operator in an angular momentum
basis: $\trme{J'}{\Jvec}{J}=\delta_{J'J}[J(J+1)]^{1/2}$.  It is nonvanishing
only between the $T=1$ states, for which we have $\trme{1}{2\vec{T}}{1}=2\sqrt{2}$ in the convention of Rose.  Again, to
convert to the normalization
convention of Edmonds~\cite{edmonds1960:am}, this matrix element may be multiplied by
$(2T'+1)^{1/2}$, \textit{i.e.}, $\sqrt{3}$.


\section{Harmonic oscillator ladder operators: Single-particle, relative, and intrinsic}
\label{sec-app-ladder}

\subsection{Ladder operators in three dimensions}
\label{sec-app-ladder-defn}

Several of the intrinsic operators considered in this work are more naturally
represented in terms of oscillator creation and annihilation, or ladder,
operators than directly in terms of the coordinate and momentum operators.  After reviewing the basic definitions and relations, we
therefore lay out properties of the ladder operators defined on the two-body
relative (Sec.~\ref{sec-relative}) and $A$-body intrinsic
(Sec.~\ref{sec-intrinsic}) degrees of freedom, as well as their complementary
c.m.\ degrees of freedom.  Care must be taken with conventional factors arising
in the definitions of coordinates and momenta, as in
Secs.~\ref{sec-intrinsic}--\ref{sec-relative}.

First, let us recall the ladder operators defined for a three-dimensional isotropic
harmonic oscillator~\cite{moshinsky1996:oscillator,rowe2010:rowanwood}, in terms
of the coordinate vector $\xvec$ and conjugate momentum vector
$\vec{p}=\hbar\kvec$.  The oscillator problem is defined by the Hamiltonian
\begin{equation}
\label{eqn-ho-hamiltonian-messy}
H=\frac{\hbar^2}{2m}\kvec^2+\frac{m\omega^2}{2}\xvec^2,
\end{equation}
with mass $m$ and oscillator frequency $\omega$ as parameters.  Although this
expression in~(\ref{eqn-ho-hamiltonian-messy}) is perhaps the most familiar form
for the Hamiltonian, it may be parametrized instead in terms of the
\textit{oscillator length} $b=[\hbar/(m\omega)]^{1/2}$, which allows the
Hamiltonian to be written more symmetrically in the coordinate and momentum as
\begin{equation}
\label{eqn-ho-hamiltonian-b}
H=\frac{\hbar\omega}{2}\bigl(
b^2\kvec^2+b^{-2}\xvec^2
\bigr),
\end{equation}
Thus, $\xvec$, with dimensions of length, and $\kvec$, with dimensions of
inverse length, are each scaled by the appropriate power of $b$ to make the
quantity in parentheses dimensionless.  The oscillator length is also
significant in that it determines the overall length scale, or dilation, of the
eigenfunctions obtained from this oscillator
Hamiltonian~(\ref{eqn-ho-hamiltonian-b})~\cite{suhonen2007:nucleons-nucleus},
but here we are primarily concerned not with the eigenfunctions but with the
operators.

The oscillator Hamiltonian~(\ref{eqn-ho-hamiltonian-b})
may then be reexpressed in terms of the oscillator creation operator $\cvec^\dagger$ and
annihilation operator $\cvec$, as $H=\hbar\omega(\cvec^\dagger\cdot\cvec+3/2)$.
The traditional expression for the creation operator is
\begin{equation}
\label{eqn-raising-messy}
\cvec^\dagger=\biggl(\frac{m\omega}{2\hbar}\biggr)^{1/2}\xvec-i \biggl(\frac{\hbar}{2 m\omega}\biggr)^{1/2}\kvec.
\end{equation}
The lowering operator $\cvec=(\cvec^\dagger)^\dagger$ is defined by taking the
adjoint of $\cvec^\dagger$, that is, more precisely, conjugating each Cartesian
component.  As with the Hamiltonian, these operators may be reparametrized in terms of the oscillator
length to yield more symmetric expressions
\begin{equation}
\label{eqn-ladder-b}
\cvec^\dagger=\frac{1}{\sqrt2}\bigl(b^{-1}\xvec-i b \kvec \bigr)
\quad
\cvec=\frac{1}{\sqrt2}\bigl(b^{-1}\xvec+i b \kvec \bigr).
\end{equation}
Inverting for the coordinate and momentum equivalently yields
\begin{equation}
\label{eqn-ladder-inverse-b}
b^{-1}\xvec=\frac{1}{\sqrt2}\bigl(\cvec+\cvec^\dagger\bigr)
\quad
b\kvec=-\frac{i}{\sqrt2}\bigl(\cvec-\cvec^\dagger\bigr).
\end{equation}
The total number of oscillator excitations (along all Cartesian axes) is then
counted by the oscillator number operator $N=\cvec^\dagger\cdot\cvec=\sum_r
c_r^\dagger c_r$ ($r=1,2,3$), in terms of which the Hamiltonian operator
in~(\ref{eqn-ho-hamiltonian-b}) is simply $H=\hbar\omega(N+3/2)$.

Recall the canonical commutation properties, for the Cartesian components, which will be useful in
understanding the relations of intrinsic, relative, and c.m.\ oscillator ladder
operators below.  Since the
coordinates and momenta obey canonical commutation relations
$[x_r,k_s]=\delta_{rs}i$ (with $[x_r,x_s]=[k_r,k_s]=0$), it follows by~(\ref{eqn-ladder-b}) that the Cartesian components of $\cvec$ and
$\cvec^\dagger$ obey canonical commutation relations
$[c_r,c^\dagger_s]=\delta_{rs}$ (with $[c_r,c_s]=[c^\dagger_r,c^\dagger_s]=0$).

The canonical commutators for vector operators can be more concisely expressed
if we consider the spherical tensor coupled
commutator~\cite{french1966:multipole,varshalovich1988:am,chen1993:wick-coupled}.
For two spherical
tensors of angular momentum (or rank) $a$ and $b$, respectively, the coupled
commutator of rank $c$ is the spherical tensor defined by
\begin{equation}
\label{eqn-commutator-coupled}
[A_a,B_b]_c=(A_a\times B_b)_c-(-)^{c-a-b}(B_b\times A_a)_c
\end{equation}
(see, \textit{e.g.}, Appendix of Ref.~\cite{caprio2011:pairalg} for a review).
From the above commutators for the Cartesian components of $\xvec$ and $\kvec$, and similarly of $\cvec$ and $\cvec^\dagger$,
we obtain coupled commutators
\begin{equation}
  \label{eqn-commutator-coupled-canonical}
        [\xvec,\kvec]_{L_0}=-\sqrt3i\delta_{L_0,0}
        \qquad
            [\cvec,\cvec^\dagger]_{L_0}=-\sqrt3\delta_{L_0,0},
\end{equation}
while
$[\xvec,\xvec]_{L_0}=[\kvec,\kvec]_{L_0}=0$ and
$[\cvec,\cvec]_{L_0}=[\cvec^\dagger,\cvec^\dagger]_{L_0}=0$.

Here we are representing the annihilation operator as a spherical tensor by
taking the rank-$1$ covariant spherical tensor $c_1$ obtained directly from the
Cartesian vector $\cvec$ via~(\ref{eqn-vector-spherical}).  However, care must
be taken when comparing with the literature. It is also common to start from
$\cvec^\dagger$, take the corresponding covariant spherical tensor
$c^\dagger_1$, and then obtain the covariant adjoint tensor
$\widetilde{(\smash{c^\dagger_1})}^\dagger$ (see, \textit{e.g.}, Sec.~4.8 of
Ref.~\cite{brink1994:am} or Sec.~A.3 of Ref.~\cite{rowe2010:rowanwood} for the
covariant adjoint of a spherical tensor).  The resulting tensor is commonly
denoted by $\tilde{c}_1\equiv \widetilde{(\smash{c^\dagger_1})}^\dagger$,
reflecting a notional but notationally dubious cancellation of the two dagger
symbols (despite their substantially different definitions).

The relation between $\cvec$ (or, rather, the corresponding covariant spherical
tensor $c_1$) and $\tilde{c}_1$ depends upon the convention in use for the
covariant adjoint.  If the covariant adjoint of a tensor $T_J$ is defined as
(\textit{e.g.}, Refs.~\cite{rowe2010:rowanwood,iachello2015:liealg})
\begin{equation}
\label{eqn-covariant-adjoint}
\tilde{T}^\dagger_{JM}=(-)^{J-M}T^\dagger_{J,-M}
\end{equation}
or as $\tilde{T}^\dagger_{JM}=(-)^{J+M}T^\dagger_{J,-M}$ (\textit{e.g.},
Ref.~\cite{suhonen2007:nucleons-nucleus}), then we have $\tilde{c}_1=-c_1$.  However,
for spherical tensors of integer rank $L$, it is also common (see Sec.~A.6 of
Ref.~\cite{rowe2010:rowanwood}) to take the alternative definition, modeled on
the conjugation property of the spherical harmonics,
\begin{equation}
\label{eqn-covariant-adjoint-spherical-harmonic}
\tilde{T}^\dagger_{LM}=(-)^{M}T^\dagger_{L,-M}.
\end{equation}
In this case, we have $\tilde{c}_1=+c_1$.

In particular, the expression
$[\tilde{A}_a,B_b^\dagger]_c=\hat{a}\delta_{AB}\delta_{c,0}$ for the generic
canonical commutator for bosonic ladder operators, as found in~(10) of
Ref.~\cite{chen1993:wick-coupled} or~(A.8) of Ref.~\cite{caprio2011:pairalg},
gives $[\tilde{c}_1,c^\dagger_1]_{L_0}=+\sqrt{3}\delta_{L_0,0}$.  This may be
reconciled with the opposite sign appearing on the right hand side of the
expression for $[\cvec,\cvec^\dagger]_{L_0}$ given above
in~(\ref{eqn-commutator-coupled-canonical}) by noting that the result of
Refs.~\cite{chen1993:wick-coupled,caprio2011:pairalg} is derived in terms of a
covariant adjoint defined under a phase convention in which $c_1=-\tilde{c}_1$.
On the other hand, the sign of the $\grpsu{3}$ generator expression
in~(\ref{eqn-sp3r-generators-su3-coupled}) is consistent with~(3) of
Ref.~\cite{escher2002:pds-symplectic}, as this is obtained under the phase
convention in which $c_1=+\tilde{c}_1$.

\subsection{Relative ladder operators (two-body system)}
\label{sec-app-ladder-rel}

For the two-particle system (Sec.~\ref{sec-relative}),
relative and c.m.\ ladder operators are obtained by a unitary change of basis on
the bosonic creation operators, from the single-particle ladder operators to the
difference and sum operators, as
\begin{equation}
\label{eqn-ladder-rel-cm}
\cvecrel^\dagger=\frac{1}{\sqrt2}\bigl(\cvec^\dagger_1-\cvec^\dagger_2 \bigr)
\quad
\cveccm^\dagger =\frac{1}{\sqrt2}\bigl(\cvec^\dagger_1+\cvec^\dagger_2 \bigr).
\end{equation}
Such a unitary change of basis automatically preserves the canonical commutation
relations (\textit{e.g.}, Ref.~\cite{caprio2005:coherent}).  That is, starting
from the canonical commutators for the individual particles,
$[\cvec_i,\cvec^\dagger_j]_{L_0}=-\sqrt{3}\delta_{ij}\delta_{L_0,0}$ ($i,j=1,2$), \textit{etc.}, we have
$[\cvecrel,\cvecrel^\dagger]_{L_0}=[\cveccm,\cveccm^\dagger]_{L_0}=-\sqrt{3}\delta_{L_0,0}$ and
$[\cvecrel,\cveccm^\dagger]_{L_0}=0$, \textit{etc.}

The form of the total number operator is also preserved under a unitary change
of basis on the bosonic opeartors, that is,
\begin{math}
N
=
\cvec^\dagger_1\cdot\cvec_1 + \cvec^\dagger_2\cdot\cvec_2
=
\cvecrel^\dagger\cdot\cvecrel+\cveccm^\dagger\cdot\cveccm.
\end{math}
The number operator may thus be decomposed into mutually-commuting relative and
c.m.\ contributions as $N=\Nrel+\Ncm$, with
$\Nrel=\cvecrel^\dagger\cdot\cvecrel$ and $\Ncm=\cveccm^\dagger\cdot\cveccm$.

However, care must be taken in identifying the relevant oscillator lengths,
under the conventions chosen for the relative and c.m.\ coordinates.  (Failing
to properly do so, when numerically evaluating integrals for relative two-body
matrix elements, for instance, leads to use of an erroneous value for the
oscillator length parameter in the relative harmonic oscillator basis functions,
and thus erroneous values for the matrix elements.)  We recognize
\begin{equation}
\begin{aligned}
\label{eqn-raising-cm-coordinates}
\cvecrel^\dagger&=\frac{1}{\sqrt2}\bigl(\brel^{-1}\xvecrel-i \brel \kvecrel \bigr)
&\ifproofpre{}{\quad}
\brel&=\Bstack{\sqrt2}{1}b
\\
\cveccm^\dagger&=\frac{1}{\sqrt2}\bigl(\bcm^{-1}\xveccm-i \bcm \kveccm \bigr)
&\ifproofpre{}{\quad}
\bcm&=\Bstack{\tfrac1{\sqrt2}}{1}b,
\end{aligned}
\end{equation}
with the upper coefficient (in braces) applying under the mechanics convention
and the lower coefficient applying under the symmetric convention for the
relative-c.m.\ coordinates, as in~(\ref{eqn-coords-rcm}).  This result is
obtained by reexpressing the relative-c.m.\ ladder operators
of~(\ref{eqn-ladder-rel-cm}) in terms of single-particle coordinates and the
oscillator length $b$ for the single-particle problem,
via~(\ref{eqn-ladder-b}), and then recognizing the relative-c.m.\ coordinates
and momenta as defined in~(\ref{eqn-coords-rcm}).

\subsection{Intrinsic ladder operators (many-body system)}
\label{sec-app-ladder-intrinsic}

For the $A$-particle system (Sec.~\ref{sec-intrinsic}), we start from the oscillator creation operators $\cvec^\dagger_i$ ($i=1,\ldots,A$) for the single-particle degrees of freedom,
\begin{equation}
\label{eqn-raising-sp}
\cvec^\dagger_i=\frac{1}{\sqrt2}\bigl(b^{-1}\xvec_i-i b \kvec_i \bigr).
\end{equation}
Then intrinsic creation operators $\cvec^{\prime\dagger}_i$ are
obtained by applying the substitutions $\xvec_i\rightarrow \xvec'_i$ and
$\kvec_i\rightarrow\kvec'_i$ in~(\ref{eqn-coords-intrinsic}), while the
c.m.\ creation operator $\cveccm^{\dagger}$ is obtained from the sum of the single-particle
creation operators:
\begin{equation}
\begin{aligned}
\label{eqn-ladder-intrinsic}
\cvec^{\prime\dagger}_i&=\cvec^\dagger_i-\frac{1}{A}\sum_j \cvec^\dagger_j&
\quad
\cveccm^{\dagger}&=\frac{1}{\sqrt{A}}\sum_i \cvec^\dagger_i.
\end{aligned}
\end{equation}

The definition of the c.m.\ coordinate and momentum
in~(\ref{eqn-coords-intrinsic}) gives the canonical commutatation relations for
the c.m.\ degree of freedom, \textit{i.e.},
$[\xveccm,\kveccm]_{L_0}=-\sqrt3i\delta_{L_0,0}$, with all others vanishing.
Thus, the above definition for $\cveccm^{\dagger}$ likewise yields canonical
commutators for the ladder operators, \textit{i.e.},
$[\cveccm,\cveccm^\dagger]_{L_0}=-\sqrt3\delta_{L_0,0}$, with all others
vanishing.  We may thus meaningfully define a c.m.\ number operator $\Ncm=
\cveccm^\dagger\cdot\cveccm$, with the usual properties for a number operator.
However, the oscillator length $\bcm$ for the c.m.\ degree of freedom is not
necessarily that associated with the single-particle oscillators.  Rather, we
identify
\begin{equation}
\label{eqn-raising-cm}
\cveccm^\dagger=\frac{1}{\sqrt2}\bigl(\bcm^{-1}\xveccm-i \bcm \kveccm \bigr)
\quad
\bcm=\Bstack{\ifproofpre{\tfrac}{\sfrac}{1}{\sqrt{A}}}{1}b,
\end{equation}
with the upper coefficient (in braces) applying under the mechanics convention
and the lower coefficient applying under the symmetric convention for the
many-body c.m.\ coordinate, as in~(\ref{eqn-coords-intrinsic}).  This result is
obtained by reexpressing the c.m.\ ladder operator $\cveccm^\dagger$
of~(\ref{eqn-ladder-intrinsic}) in terms of single-particle coordinates and the
oscillator length $b$ for the single-particle problem,
via~(\ref{eqn-raising-sp}), and then recognizing the c.m.\ coordinate $\xveccm$
and momentum $\kveccm$ defined in~(\ref{eqn-coords-intrinsic}).  The
c.m.\ oscillator length $\bcm$ in~(\ref{eqn-raising-cm}) defines the length
scale for the zero-point motion of the c.m.\ degree of freedom, which plays an
important role in oscillator-basis calculations~\cite{elliott1955:com-shell},
\textit{e.g.}, when the explicit form of this zero-point motion must be used to
apply corrections in NCCI calculations~\cite{cockrell2012:li-ncfc}.

However, the overcomplete set of intrinsic coordinates and momenta defined
by~(\ref{eqn-coords-intrinsic}) do not obey canonical commutation relations.  In
fact, it may be verified that
$[\xvec'_i,\kvec'_j]_{L_0}=-\sqrt3i\delta_{L_0,0}(\delta_{ij}-1/A)$.  Likewise,
the intrinsic ladder operators do not obey canonical commutation relations, but,
rather, have
$[\cvec'_i,\cvec^{\prime\dagger}_j]_{L_0}=-\sqrt3\delta_{L_0,0}(\delta_{ij}-1/A)$.
This noncanonicality may be understood since the overcomplete set of ladder operators defined
in~(\ref{eqn-ladder-intrinsic}) do not provide a unitary change of basis on
the bosonic creation operators $\cvec^\dagger_i$ defined
in~(\ref{eqn-raising-sp}).  Nonetheless, the intrinsic ladder operators do
commute with the c.m.\ ladder operators, \textit{e.g.},
$[\cvec'_i,\cveccm^\dagger]_{L_0}=0$.  Moreover, it may be verified that the
total number operator $N=\sum_i \cvec^\dagger_i\cdot\cvec_i$ separates into
mutually-commuting intrinsic and c.m.\ contributions as $N=N'+\Ncm$, where
$N'=\sum_i \cvec^{\prime\dagger}_i\cdot\cvec'_i$.


\providecommand{\newblock}{}



\end{document}